\documentclass[a4paper]{scrartcl}
\usepackage[ansinew]{inputenc}
\usepackage{amsmath,amssymb,amstext,color,bm,verbatim,graphicx,float}
\usepackage[margin=0.9in,bottom=1in,top=1in]{geometry}
\usepackage{fancyhdr}
\usepackage{lastpage}
\usepackage{cancel}
\usepackage{booktabs}
\usepackage{enumitem}
\usepackage{graphicx}
\usepackage{xcolor}
\usepackage{times}
\makeatletter

\begin{document}
\begin{center}
\huge{\textbf{Selective excitation of work-generating cycles in 
\\ nonreciprocal living solids}} 
\end{center}
\vspace{0.2cm}

\normalsize
Yu-Chen Chao$^{1,6}$, Shreyas Gokhale$^{1,6}$, Lisa Lin$^{1,6}$, Alasdair Hastewell$^{2,6}$, 
\\Alexandru Bacanu$^3$, Yuchao Chen$^1$, Junang Li$^{1,4}$, Jinghui Liu$^{1,5}$, Hyunseok Lee$^1$, 
\\J\"{o}rn Dunkel$^2$, Nikta Fakhri$^{1,7}$

\vspace{-0.25cm}

\begin{itemize}[leftmargin=0.15in]\small
\item[$^1$] Department of Physics, Massachusetts Institute of Technology, Cambridge, MA, USA \vspace{-0.3cm}
\item[$^2$] Department of Mathematics, Massachusetts Institute of Technology, Cambridge, MA, USA \vspace{-0.3cm}
\item[$^3$] Department of Molecular and Cellular Biology, Harvard University, Cambridge, MA, USA \vspace{-0.3cm}
\item[$^4$] Center for the Physics of Biological Function, Princeton University, Princeton, NJ, USA \vspace{-0.3cm}
\item[$^5$] Center for Systems Biology Dresden, Dresden, Germany \vspace{-0.3cm}
\item[$^6$] These authors contributed equally and are joint first authors. \vspace{-0.3cm}
\item[$^7$] Corresponding author: fakhri@mit.edu
\end{itemize}

\vspace{1cm}
\normalsize
\section*{Abstract}

Emergent nonreciprocity in active matter drives the formation of self-organized states that transcend the behaviors of equilibrium systems. Integrating experiments, theory and simulations, we demonstrate that active solids composed of living starfish embryos spontaneously transition between stable fluctuating and oscillatory steady states. The nonequilibrium steady states arise from two distinct chiral symmetry breaking mechanisms at the microscopic scale: the spinning of individual embryos resulting in a macroscopic odd elastic response, and the precession of their rotation axis, leading to active gyroelasticity. In the oscillatory state, we observe long-wavelength optical vibrational modes that can be excited through mechanical perturbations. Strikingly, these excitable nonreciprocal solids exhibit nonequilibrium work generation without cycling protocols, due to coupled vibrational modes. Our work introduces a novel class of tunable nonequilibrium processes, offering a framework for designing and controlling soft robotic swarms and adaptive active materials, while opening new possibilities for harnessing nonreciprocal interactions in engineered systems.

\newpage

Symmetries and conservation laws  impose stringent constraints on material properties, often limiting their functional capabilities. For instance, time-reversal symmetry and energy conservation prevent extracting work from strain cycles in conventional linear elastic solids \cite{landau1986theory}. Such constraints, however, can be overcome in systems composed of active microscopic building blocks that consume free energy to generate motion \cite{needleman2017active, bowick2022symmetry}. Recent research has exploited the broken symmetries of active particles and their collective dynamics to reveal a spectrum of nonequilibrium phenomena and unconventional material properties. These discoveries include topological modes in gyroscopic materials \cite{nash2015topological} and active fluids \cite{souslov2017topological}, active elastocapillarity \cite{binysh2022active}, odd viscosity \cite{banerjee2017odd, de2024pattern}, odd viscoelasticity in colloidal fluids \cite{soni2019odd, bililign2022motile}, odd elasticity in active solids \cite{scheibner2020odd, tan2022odd, shankar2024active}, collective actuation of elastic modes \cite{baconnier2022selective}, the non-Hermitian skin effect \cite{scheibner2020non, ghatak2020observation}, and nonreciprocal phase transitions \cite{fruchart2021non, dinelli2023non, ceron2023programmable, packard2022non, you2020nonreciprocity, loos2023long, hanai2024nonreciprocal}. Such advancements have inspired the design of materials with unconventional mechanical properties, such as nonreciprocal robotic metamaterials \cite{brandenbourger2019non, veenstra2024soliton}, solids with odd micropolar elasticity \cite{chen2021realization}, and odd wheels that spontaneously roll uphill \cite{brandenbourger2021limit}.
\\

In many of these systems, the microscopic breaking of parity and time-reversal symmetries results in nonreciprocal couplings between components of displacement or velocity fields on a macroscopic scale \cite{fruchart2023odd}. This rich underlying physics and versatile phenomenology place nonreciprocal materials at the forefront of research on active and intelligent matter \cite{veenstra2024soliton, nassar2020nonreciprocity, wang2023mechanical, wang2023non}. Yet, we are just beginning to understand the full potential of these systems. Two major questions arise regarding the connection between the macroscopic properties of nonreciprocal materials and the microscopic mechanisms of symmetry breaking: Do distinct mechanisms of symmetry breaking produce distinct nonequilibrium states in nonreciprocal materials? If so, can we control the symmetry-breaking mechanism to transition between these distinct states?
\\

We provide affirmative answers to both questions using the paradigmatic example of living chiral crystals (LCCs) of starfish embryos. Specifically, we show that individual embryos bound near the air-water interface exhibit two distinct mechanisms of parity breaking: rotation, or spin, about their anterior-posterior (AP) axis; and tilt and precession of this axis about the vertical. The symmetry breaking at the microscopic scale contribute to macroscopic behaviors, where LCCs can spontaneously switch between two stable states - one characterized by active fluctuations and the other by self-sustained oscillations. 
We introduce a novel, data-driven method to extract the dispersion relation by decomposing embryo displacements into spatial vibrational modes, unveiling an emergent optical branch that is observable only in the oscillatory state. Analysis of the coupling between these vibrational modes uncovers different phase space cycles generated by optical and acoustic mode dynamics, indicating two distinct oscillation mechanisms in LCCs. A continuum model combining linear odd elasticity with tilt and precession dynamics captures these experimental observations, linking the optical branch associated with the oscillatory state of the cluster to the precession of the tilted embryos. Notably, we demonstrate that the oscillatory state can be selectively and reproducibly excited from the fluctuating state using controlled mechanical perturbations. Furthermore, by estimating entropy production rates, we quantify the work generated by these selectively excited LCCs, revealing the non-equilibrium nature of the system and the potential for sustained oscillatory behavior to drive work without external cyclic forcing. Our findings advance the development of excitable nonreciprocal solids into functional active materials.

\subsection*{LCCs exhibit two stable states}
\begin{figure}[h!]
  \centering\includegraphics[width=0.9\linewidth]{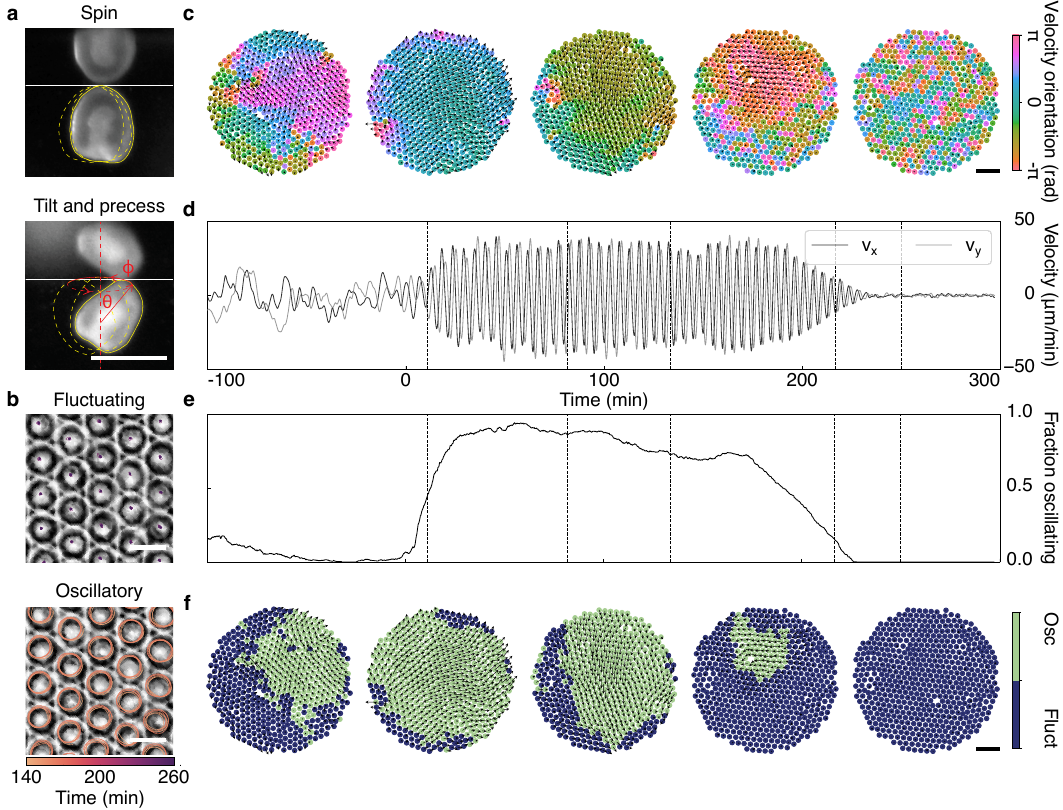}
    \caption{\textbf{Experimental observation of two stable states of LCCs arising from distinct mechanisms of chiral symmetry breaking at the embryo level.} \textbf{(a)} Individual embryos can spin with their body axes perpendicular to the air-water interface (upper), or precess with body axes that make a finite angle $\theta$ with the vertical at rate $\dot \phi$ (lower). Dashed outlines indicate embryo positions at earlier time points, with all outlines spaced one second apart. \textbf{(b)} Zoomed-in snapshot of a living  chiral crystal exhibiting both stable fluctuating (upper) and oscillatory (lower) states, with individual embryo trajectories overlaid. \textbf{(c)} Velocity snapshots in a cropped center region of the system, with arrows indicating instantaneous velocity vectors and color indicating velocity orientation. \textbf{(f)} Binarized map indicating spatial growth and decay of oscillations in the same region as \textbf{(c)} (see section 3.2 in the supplementary text). Selected time points in \textbf{(c)} and \textbf{(f)} correspond to the vertical dashed lines in \textbf{(d)} and \textbf{(e)}. Time series of \textbf{(d)} velocity and \textbf{(e)} fraction of oscillating embryos, averaged spatially over the region indicated in \textbf{(c)} and \textbf{(f)}, illustrate transitions between fluctuating and oscillatory states. Scale bars in \textbf{(a)}, \textbf{(b)} correspond to 250 $\mu$m; those in \textbf{(c)}, \textbf{(f)} correspond to 1 mm.}
  \label{fig:2states}
\end{figure}

Starfish embryos can self-assemble into LCCs near the air-water interface \cite{tan2022odd} due to hydrodynamic interactions \cite{ishikawa2006hydrodynamic, ruhle2022gyrotactic} similar to those observed in other swimming microorganisms \cite{drescher2009dancing}. When viewed laterally, individual embryos form two distinct bound states: i) an upright state, where embryos spin about an AP axis perpendicular to the air-water interface (i.e., the vertical direction, see Fig. \ref{fig:2states}a, top panel); and ii) a tilted state, where the AP axis forms a finite angle with the vertical, and precesses around it (Fig. \ref{fig:2states}a, bottom panel). In this tilted state, the tip of the embryo traces circular orbits at the air-water interface. These two bound states likely result from the two modes of locomotion - rectilinear and helical - observed for freely swimming embryos within the bulk fluid (section 1.2.1 of the supplementary text). To assess whether these microscopic bound states influence macroscopic LCC dynamics, we conducted video microscopy experiments on LCCs at the air-water interface (section 1.3 of the supplementary text). We selected the developmental time of embryos to ensure that LCCs remained stable for several hours without dissolving \cite{tan2022odd}. In many experiments, we observed that LCCs could persist in either a ``fluctuating state'', wherein embryo positions only exhibit small fluctuations due to the active hydrodynamic flows generated by their cilia (Fig. \ref{fig:2states}b, top panel), or an ``oscillatory state'' (Fig. \ref{fig:2states}b, bottom panel) characterized by stable velocity oscillations.
\\

In typical experiments, these velocity oscillations spontaneously emerge, growing in amplitude until they reach a steady-state value. Remarkably, they can persist for several hours before decaying, leading the LCC to revert to its non-oscillatory fluctuating state (Fig. \ref{fig:2states}c-d and section 3.2 of the supplementary text). This growth in oscillation amplitude is accompanied by an initial expansion of the oscillating region, which eventually encompasses the bulk of the system at steady state. Subsequently, the oscillating region contracts and eventually vanishes as the system returns to the fluctuating state (Fig. \ref{fig:2states}e-f and section 3.2 of the supplementary text). The bistability observed between fluctuating and oscillatory states here contrasts with our previous work \cite{tan2022odd}, where oscillations exhibited more transient growth and decay.
\\

We propose that the tilted bound state of embryos (Fig. \ref{fig:2states}a, bottom panel) is a plausible mechanism for collective, sustained oscillations. Collective precession of embryos about the axis perpendicular to the air-water interface, when projected onto the imaging surface, would manifest as planar orientation-dependent self-propulsion of the embryo centroids (as in Fig. \ref{fig:2states}b, bottom panel). Thus, we hypothesize that the transition between fluctuating and oscillatory states in the LCC results from a collective switching of the embryos from the upright spinning state to the tilted precessing state.

\subsection*{Mode decomposition reveals acoustic and optical vibrational modes}

\begin{figure}[h!]
  \centering
  \includegraphics[width=\linewidth]{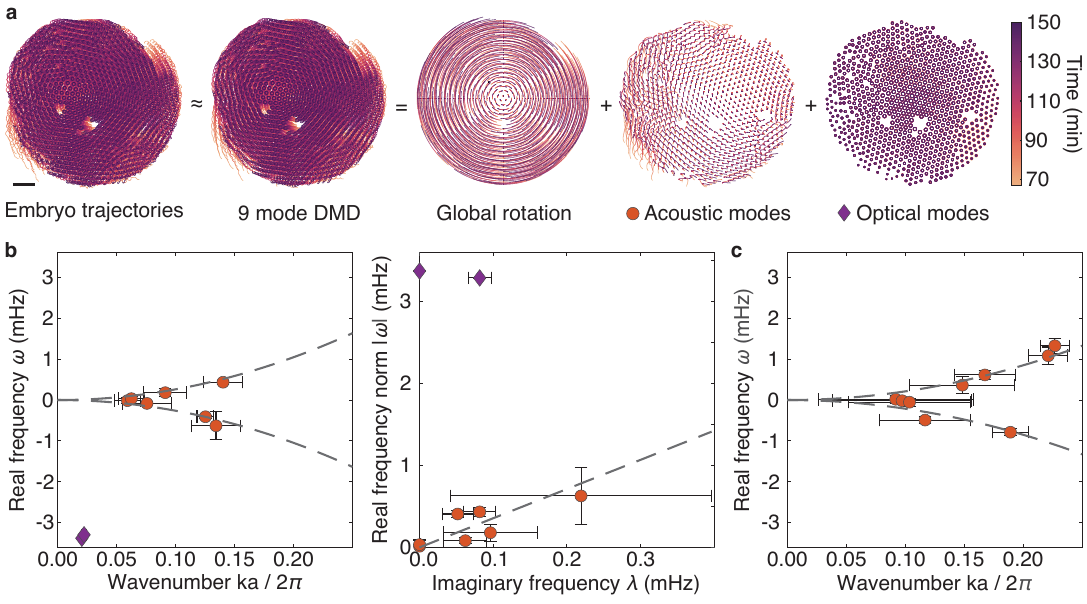}
    \caption{\textbf{Vibrational mode decomposition and dispersion relations for fluctuating and oscillatory states}. \textbf{(a)} Living chiral crystal dynamics can be decomposed into global rotation, acoustic modes, and optical modes; scale bar corresponds to 1 mm. \textbf{(b, left)} During the oscillatory state, the dispersion relation displays both a quadratic acoustic branch and a flat optical branch. \textbf{(b, right)} The optical branch displays low decay. Ratios of oscillation frequency to decay rate in the acoustic modes agree with the functional form of the dispersion predicted by odd elasticity, as do their nearly identical quadratic fits in both \textbf{(b)} and \textbf{(c)} (black dashed lines); the mode corresponding to global rotation is not shown. \textbf{(c)} The fluctuating state's dispersion relation only displays signatures of the quadratic acoustic band. }
  \label{fig:DispersionRelation}
\end{figure}

To gain a deeper understanding of LCC dynamics in the fluctuating and oscillatory states, we analyzed the displacement field in terms of its constituent spatiotemporal vibrational modes, obtained via Dynamic Mode Decomposition (DMD). Strikingly, the displacement dynamics of the oscillatory and fluctuating states can be accurately reconstructed with a low rank representation consisting of less than ten modes (see section 3.6.2 of the supplementary text for quantification of quality of reconstruction). Fig. \ref{fig:DispersionRelation}a illustrates the decomposition of oscillatory state dynamics into nine dynamical modes in addition to global LCC rotation. The global rotation originates from the pre-stress induced by transverse force exchange between spinning embryos \cite{tan2022odd}. Seven of the modes correspond to long wavelength, low frequency modulations of the displacement reminiscent of acoustic vibrations in elastic solids. The remaining two modes correspond to highly localized, high frequency circular motion about the embryo centroid, consistent with precession dynamics. 
\\

The remarkable accuracy of our linear DMD analysis in reconstructing LCC dynamics suggests that, while nonlinear effects are necessary to explain switching between fluctuating and oscillatory states, linear dynamics dominate within each stable state. The results of the mode decomposition analysis can be concisely summarized in terms of dispersion relations $\omega(q)$ for LCC lattice vibrations (Fig. \ref{fig:DispersionRelation}b-c; see section 3.6 of the supplementary text for details). Crucially, DMD allows us to extract the dominant temporal frequencies $\omega$ within our system without enforcing orthogonality of dynamical mode vectors. In general, the frequencies $\omega$ are complex numbers whose real part yields the oscillation period, and imaginary part yields the decay, or dissipation, rate of the mode. Having extracted $\omega$, we represent the space-dependent DMD mode coefficients as linear combinations of the orthonormal graph Laplacian eigenmodes, obtained from triangulation of the ordered LCC, to determine the DMD modes' associated spatial frequencies $q$. 
\\

In contrast to the acoustic modes, the localized modes shown in Fig. \ref{fig:DispersionRelation}a exhibit a finite frequency even in the long wavelength limit ($q\to 0$), reminiscent of optical modes in lattice vibrations \cite{Ashcroft76}. We further observe that these long-wavelength optical modes have low dissipation rates (Fig. \ref{fig:DispersionRelation}b, right) and can therefore sustain constant oscillation amplitudes over extended periods of time (Fig. \ref{fig:2states}d-e). The real part of the acoustic modes' temporal frequencies is nearly an order of magnitude larger than the imaginary part, even in the fluctuating LCC (Fig. \ref{fig:DispersionRelation}b, right, and section 3.6.1 of the supplementary text). This finding aligns with the ratio between odd and even elastic moduli of LCCs estimated from strain fields \cite{tan2022odd} and confirms that LCCs are odd elastic solids in the active wave regime \cite{scheibner2020odd} irrespective of whether they occupy a fluctuating or oscillatory stable state. The optical modes vanish entirely in the fluctuating LCC (Fig. \ref{fig:DispersionRelation}c), further suggesting that bistability in embryo states may underlie bistability in LCC states. In this context, the acoustic modes represent the odd elastic response of LCCs, which originates microscopically from the embryos' intrinsic spin about their body axes \cite{tan2022odd}, and is therefore present in both stable states. Conversely, the optical modes represent characteristics unique to the oscillatory state and consequently  disappear when a majority of embryos are no longer precessing.
\\

\begin{figure}[h!]
  \centering
  \includegraphics[width=\linewidth]{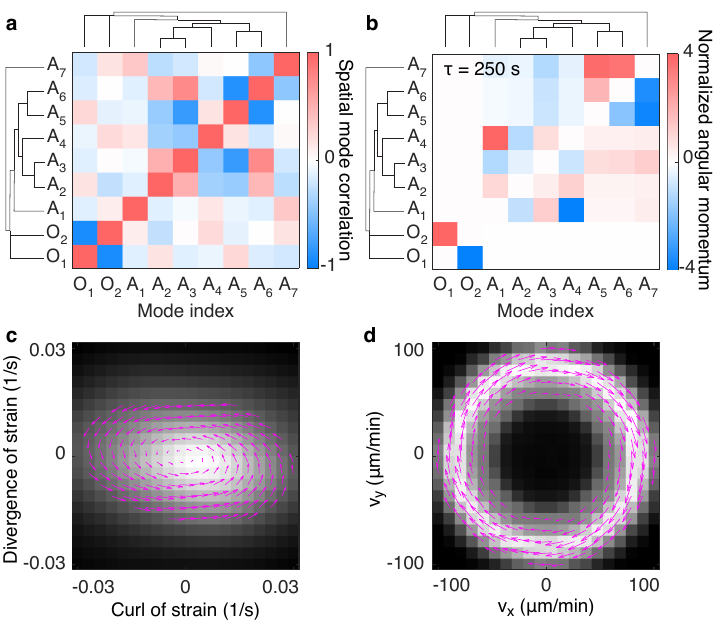}
    \caption{\textbf{Spatiotemporal mode correlations and phase space cycles.} \textbf{(a)} Spatial distance (dot product) between modes; modes have been hierarchically clustered to visualize similar groupings of mode behaviors. `A' denotes acoustic modes and `O' denotes optical modes. \textbf{(b)} Time-asymmetric correlations between hierarchically clustered modes for lag time $\tau = 250s$ (see section 3.7.3 of the supplementary text). \textbf{(c)} A clear strain cycle can be visualized from a displacement reconstruction using two acoustic modes ($A_1$ and $A_4$), as predicted by odd elasticity. \textbf{(d)} A clear velocity cycle, absent from the acoustic modes (see section 3.8 of the supplementary text), can be found from displacement reconstructions using the two optical modes.} 
  \label{fig:angmom}
\end{figure}

We further explore the effects of these two mode groupings on LCC dynamics through the analysis of vibrational mode couplings. Notably, the cosine distance between the DMD modes (Fig. \ref{fig:angmom}a) and time-asymmetric correlations \cite{bacanu2023inferring} (Fig.~\ref{fig:angmom}b), which systematically quantify scale-dependence of nonequilibium activity, indicate strong spatial and temporal correlations within the acoustic and optical modes individually, but little, if any, correlation between them. This relative lack of overlap between acoustic and optical modes in both space and time results in the coexistence of qualitatively distinct phase cycles. For instance, the displacement field reconstructed from a highly correlated, purely acoustic mode pair produces clear strain cycles (Fig. \ref{fig:angmom}c) but no velocity cycles (supplementary text section 3.8), as predicted by odd elasticity theory. Conversely, reconstruction from a pure optical mode pair leads to the emergence of velocity cycles (Fig. \ref{fig:angmom}d). These findings collectively point to the coexistence of two distinct mechanisms of oscillation: embryo spin, associated with odd elastic strain oscillations (acoustic modes); and precession, associated with sustained velocity oscillations in the bulk (optical modes). 

\subsection*{Continuum model explains oscillatory dynamics}

To test whether bistability between the upright spinning state (Fig. \ref{fig:2states}a, top panel) and the tilted precessing state (Fig. \ref{fig:2states}a, bottom panel) can account for our experimental observations, we developed a continuum theoretical model incorporating odd elasticity, as well as tilt and precession dynamics. Here, we provide the physical intuition behind the model, deferring a more detailed technical description to sections 2.1-2.3 of the supplementary text. Building on our previous finding that LCCs in the fluctuating state are odd elastic solids \cite{tan2022odd}, our model assumes overdamped linear odd elastic dynamics for the 2D displacement field $\mathbf{u}(\mathbf{r},t)$ \cite{scheibner2020odd}. As shown in Fig. \ref{fig:2states}b, bottom panel, tilt and precession result in orientation-dependent self-propulsion of embryos in the imaging plane, which leads to collective oscillations. Our model captures this self-propulsion by introducing the tilt and azimuthal angle fields, $\theta(\mathbf{r},t)$ and $\phi(\mathbf{r},t)$, respectively, which characterize the coarse grained embryo orientations in 3D. Furthermore, to account for bistability between fluctuating and oscillatory LCC states (Fig. \ref{fig:2states}d-e), our model allows for two macroscopic, spatially uniform, stable states: a fluctuating state with $\theta(\mathbf{r},t) = 0$, and an oscillatory state with $\theta(\mathbf{r},t) = \theta_0 > 0$. Due to the distinctive feedback between embryo displacements ($\mathbf{u}(\mathbf{r},t)$), and orientations ($\theta(\mathbf{r},t)$ and $\phi(\mathbf{r},t)$), our model can successfully reproduce several experimental observations. 
\\

\begin{figure}[h!]
  \centering
  \includegraphics[width=0.95\linewidth]{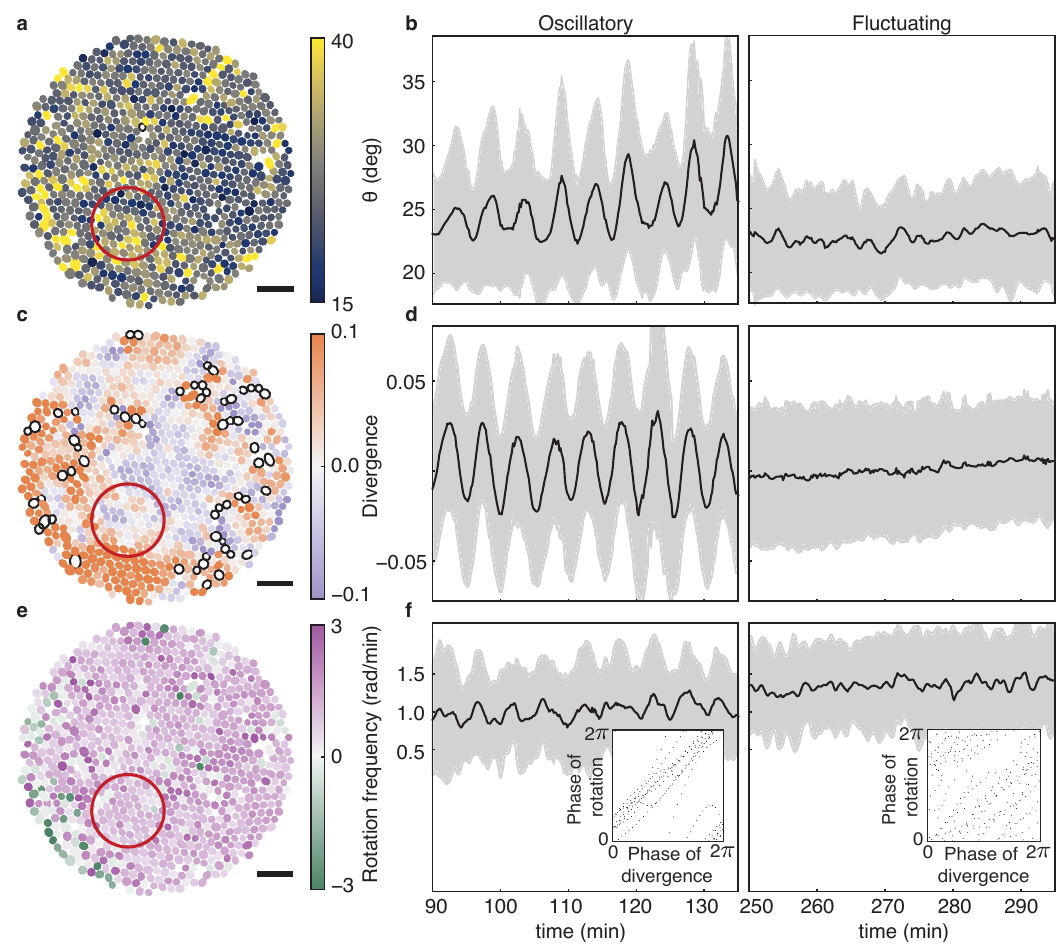}
    \caption{\textbf{Experimental verification of theoretical predictions of oscillatory dynamics.}  \textbf{(a, c, e)} Snapshots of embryos colored by tilt angle, divergence, and rotation frequency, alongside time series for the average value of each quantity within the red circle shown during the oscillating \textbf{(b, d, f left)} and fluctuating \textbf{(b, d, f right)} states. Embryos with NaN values are outlined in black (see sections 3.9 and 3.10.3 of the supplementary text for details). \textbf{(f), insets} Phase of average embryo rotation frequency plotted against average phase of divergence (see section 3.10.3 of the supplementary text for details on phase calculations.) shows that rotation frequency and divergence are in-phase while the system is undergoing long-wavelength oscillations. Scale bars in \textbf{(a, c, e)} correspond to 1 mm.}
  \label{fig:theory}
\end{figure}

Simulations of our model (eqns. S1-S4 in section 2.1 of the supplementary text) confirm the existence of stable fluctuating as well as oscillatory states across a wide range of model parameters (see section 4 of the supplementary text for details). Our theory also explains the observed spatial expansion and contraction of oscillating regions (Fig. \ref{fig:2states}c-f). In the absence of coupling between tilt and displacement, the tilt dynamics in our model are analogous to the spatial dynamics of expanding populations observed in ecological systems \cite{taylor2005allee} (See eqn. S4 and section 2.4 of the supplementary text). Specifically, in our system, a region of large tilt can `invade' a region of low tilt via an expanding `tilt wave', which manifests as an increase in the area of the oscillating region. Similarly, depending on the parameters, a high tilt state can become unfavorable, causing the oscillating region to shrink until a uniformly fluctuating state remains. Simulations of our theoretical model validate this expectation and delineate parameter regimes in which stable oscillations are permitted. Furthermore, these results remain qualitatively unaffected even in the presence of tilt-displacement coupling, with only a slight reduction in the parameter range over which oscillations can be observed. 
\\

To obtain theoretical dispersion relations, we analyzed a linearized version of our model (see section 2.5 of the supplementary text for details). In the fluctuating LCC, where $\theta = 0$ uniformly, our model reduces identically to an overdamped 2D odd elastic solid, and is characterized by a quadratic dispersion relation, $\omega(q) \propto q^2$ \cite{scheibner2020odd}. In the oscillatory state, where $\theta = \theta_0$ uniformly, the effect of tilt-induced orientation dependent self-propulsion on the displacement field leads to the emergence of two additional branches, whose dispersion is given by $\omega_{\mathrm{op}} = \pm \omega_p$, where $\omega_p$ represents the precession frequency. These branches have temporal frequencies that are finite and independent of wave vector $\mathbf{q}$ in the long wavelength limit, a hallmark feature of optical modes in lattice vibrations \cite{Ashcroft76}. The prediction of our linearized model is thus in concord with our experimental observation of optical modes (Fig. \ref{fig:DispersionRelation}b).
\\

Our continuum model also makes a number of experimentally testable predictions. For instance, our model predicts that oscillations in the divergence of the displacement field, $\partial_k u_k$, should induce oscillations in the tilt angle field $\theta(\mathbf{r},t)$ itself, due to the presence of tilt-displacement coupling (eqn. S4 in section 2.1 of the supplementary text). To test this prediction, we estimated individual embryo tilt angles from elliptical projections of the embryos on the imaging plane (see section 3.9 of the supplementary text). To compare with the coarse-grained local tilt angle $\theta(\mathbf{r},t)$, we spatially averaged individual embryo tilt angle trajectories over the spatial region enclosed by the red circle shown in Fig. \ref{fig:theory}a. As predicted, Fig. \ref{fig:theory}b shows that our extracted coarse grained local tilt angle oscillates in the oscillatory state, but only exhibits fluctuations in the fluctuating LCC, a feature corroborated across multiple experiments.
\\

Based on the coupling between the displacement field and the embryos' intrinsic rotation, or spin, our theory predicts that the divergence of the displacement field and the average rotation frequency must oscillate in phase with each other, in the oscillatory state (section 2.3 of the supplementary text). To test this prediction, we quantified individual embryo rotation frequencies by measuring the angular displacement between successive frames in the embryo frame of reference (section 3.10.2 of the supplementary text). In agreement with the model's prediction, our measurements confirm that divergence (Fig. \ref{fig:theory}c-d) and rotation frequency (Fig. \ref{fig:theory}e-f) are in-phase synchronized in the oscillatory state, but show no clear phase relationship in the fluctuating state (Fig. \ref{fig:theory}f, inset).

\subsection*{Oscillations can be selectively excited via mechanical perturbation}

Having established the existence of bistability in LCCs, we then asked whether it is possible to controllably excite the oscillatory state from the fluctuating state. To investigate this, we conducted a series of experiments in which we applied controlled mechanical perturbations to fluctuating LCCs confined in a C-shaped chamber. In these experiments, a piston attached to a syringe pump served as a movable fourth wall, allowing for uniaxial step compression over compression rates ranging from approximately $10^{-4} - 10^{-2}$ 1/s (see schematic in Fig. \ref{fig:resplot}a and section 1.3 of the supplementary text). Across approximately 80 compression steps distributed over around 25 independent experiments, we observed that sustained oscillations can not only be mechanically induced from the fluctuating state, but they can also be reproducibly excited only within a narrow band of compression rates (Fig. \ref{fig:resplot}b-d and section 3.11 of the supplementary text). At low compression rates, slow long-wavelength relaxations fail to excite oscillations (top panel of Fig. \ref{fig:resplot}b-c), while high compression rates lead to oscillations that decay rapidly due to localized plastic rearrangements (bottom panel of Fig. \ref{fig:resplot}b-c). Notably, we observe that for appropriate compression rates, oscillations can be induced repeatedly, using successive compression steps. These observations are qualitatively consistent with our linearized theory, expanded about $\theta=0$, which robustly predicts that oscillations can be selectively excited over an intermediate range of compression rates (section 2.6 of the supplementary text).

\begin{figure}[h!]
  \centering
\includegraphics[width=\linewidth]{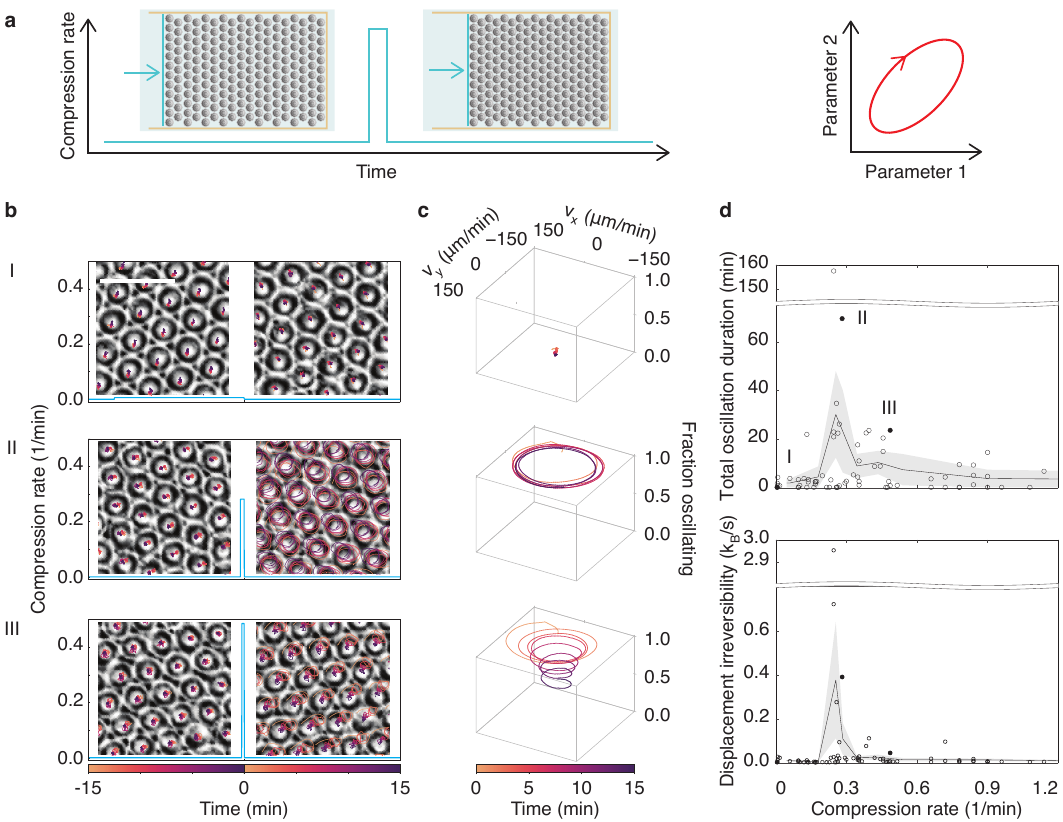}
    \caption{\textbf{Selective excitation of oscillations using mechanical compression.} \textbf{(a)} Schematic of the setup for compression experiments, as viewed from the top (left panel). We uniaxially apply a constant compression rate on the LCC between times $t_{1}$ and $t_{2}$ (middle panel). Compression-induced oscillations lead to quantifiable work-generating cycles in strain and velocity space (right panel). \textbf{(b)} Zoomed-in snapshots of an LCC before (left) and after (right) applying mechanical step compressions at slow (top), intermediate (middle), and fast (bottom) compression rates, with individual embryo trajectories overlaid; their corresponding velocity phase space trajectories are shown in \textbf{(c)}. Slow compressions generate only adiabatic displacements of embryos, while fast compressions lead to large neighbor rearrangements and damped oscillations. \textbf{(d)} Only a narrow band of intermediate compression rates lead to sustained oscillations on the order of hours (top); the same parameter regime produces a peak in the entropy production rate (bottom), signifying the largest amount of nonequilibrium activity due to mechanical compression. See section 3.12 of the supplementary text for details. Scale bars in \textbf{(b)} correspond to 500 $\mu$m.}
  \label{fig:resplot}
\end{figure}

The exploration of active materials as a means to generate work has garnered significant attention in recent years. While earlier studies have shown that work can be extracted from strain cycles in odd elastic solids \cite{scheibner2020odd, brandenbourger2021limit}, LCCs in the oscillatory state present a remarkable opportunity to harness work from displacement cycles alone, even in the absence of strain, due to the presence of optical modes (Fig. \ref{fig:DispersionRelation}b and Fig. \ref{fig:angmom}d). We observe strong signatures of irreversibility associated with these displacement cycles during selectively excited oscillations (Fig. \ref{fig:resplot}d, bottom, and section 3.12 of the supplementary text). In traditional odd elastic work cycles, work generation depends on externally applied cyclic deformations. However, switching from fluctuating to oscillatory states in LCCs allows for work generation without the need for such external interventions, with the amount of work extracted being solely determined by the system's internal parameters. The coexistence of strain cycles associated with acoustic modes and displacement, or velocity cycles associated with optical modes (Fig. \ref{fig:angmom}c-d) highlights the exciting possibility of coupling these two forms of work generation. Our quantification of vibrational mode coupling in Fig. \ref{fig:angmom}a-b underscores this potential, demonstrating the intricate dynamics at play within LCCs.
 
\subsection*{Discussion}

We have identified a fundamental mechanism by which two complementary  modes of parity breaking at the microscopic scale, active particle spinning and precession, drive emergent bistable behavior in nonreciprocal solids. This is observable as spontaneous transitions between fluctuating and oscillatory states in LLCs. Notably, our compression experiments demonstrate that the oscillatory state can be selectively excited through external mechanical perturbations. The interplay between tilt-induced self-propulsion and odd elasticity distinguishes LLCs as a unique class of excitable nonreciprocal solids, setting them apart from systems such as odd elastic \cite{scheibner2020odd}, gyroelastic \cite{fruchart2023odd}, gyro-odd \cite{gao2022non}, and elasto-active \cite{baconnier2022selective} solids. Unlike conventional gyroelastic materials, which converge to odd elastic solids under strong damping \cite{fruchart2023odd}, LCCs in the oscillatory state are active gyroelastic materials where odd elasticity and gyroelasticity can be independently controlled. 
\\

The DMD-based mode analysis framework allowed us to  directly  measure the  dispersion relations and associated vibrational mode couplings in LCCs.
The simultaneous presence of both optical and acoustic modes in LCCs gives rise to an intriguing multi-scale dynamics in which short-range excitations coexist with long-range collective behavior. In particular, the emergence of self-sustained optical vibrations within a 2D triangular monolayer-lattices is an inherently non-equilibrium phenomenon as such modes are absent in corresponding equilibrium condensed matter systems \cite{Ashcroft76}. Due to its generic mathematical formulation, the mode-decomposition framework  is broadly applicable and enables a unifying quantitative analysis of a wide range of non-equilibrium phenomena, including topological mechanics \cite{kane2014topological, sussman2016topological, meeussen2020topological} and non-Hermitian effects \cite{wang2022NHSE, librandi2021non, ghatak2020observation}, nonlinear excitations in nonreciprocal metamaterials \cite{veenstra2024soliton, Shaat2020_non, Li2018_non}, and information flow across scales in active systems \cite{de2024pattern, Coulais2017staticnon, Nicoletti2024}. 
\\

Our compression experiments demonstrate how the mode structure of non-reciprocal active solids can be leveraged to extract work from displacement cycles in a controlled and robust manner. The experimentally observed  spontaneous transitions  between fluctuating and oscillatory states in LLCs indicate a self-organized energy partitioning between the underlying  modes. This bistability highlights the potential to exploit nonlinear mode-coupling mechanisms to achieve tunable state-switching and adaptive material responses. Thus,  our experimental results and theoretical analysis establish and demonstrate the design principles for a novel class of adaptive multi-functional active materials.

\newpage
\subsection*{Acknowledgments}
This research was supported by a Sloan Foundation Grant (G-2021-16758) to N.F. and J.D., and a National Science Foundation CAREER Award (PHY-1848247) to N.F.. S.G. and H.L. acknowledge the Gordon and Betty Moore Foundation for support as Physics of Living Systems Fellows through Grant No. GBMF4513. L.L. was supported by the National Science Foundation for a Graduate Research Fellowship under Grant No. 2141064. J.L. acknowledges the support of the Center for the Physics of Biological Function (PHY-1734030). 
This research received support through Schmidt Sciences, LLC (to J.D.), the MathWorks Professorship Fund (to J.D.), and National Science Foundation Award DMR-2214021 (to J.D.). N.F. and J.D. thank the WPI-SKCM$^2$ Hiroshima University for hospitality and support. The authors acknowledge the MIT SuperCloud and Lincoln Laboratory Supercomputing Center for providing HPC resources that have contributed to the research results reported within this paper. 
\\

\bibliographystyle{unsrt}
\bibliography{references}

\end{document}


\begin{center}
\Huge{\textbf{Supplementary Information}}
\end{center}
\begin{center}

\LARGE{\textbf{Selective excitation of work-generating cycles in 
\\ nonreciprocal living solids}} 
\end{center}

\normalsize
Yu-Chen Chao$^{1,6}$, Shreyas Gokhale$^{1,6}$, Lisa Lin$^{1,6}$, Alasdair Hastewell$^{2,6}$, 
\\Alexandru Bacanu$^3$, Yuchao Chen$^1$, Junang Li$^{1,4}$, Jinghui Liu$^{1,5}$, Hyunseok Lee$^1$, 
\\J\"{o}rn Dunkel$^2$, Nikta Fakhri$^{1,7}$

\vspace{-0.25cm}

\begin{itemize}[leftmargin=0.15in]\small
\item[$^1$] Department of Physics, Massachusetts Institute of Technology, Cambridge, MA, USA \vspace{-0.3cm}
\item[$^2$] Department of Mathematics, Massachusetts Institute of Technology, Cambridge, MA, USA \vspace{-0.3cm}
\item[$^3$] Department of Molecular and Cellular Biology, Harvard University, Cambridge, MA, USA \vspace{-0.3cm}
\item[$^4$] Center for the Physics of Biological Function, Princeton University, Princeton, NJ, USA \vspace{-0.3cm}
\item[$^5$] Center for Systems Biology Dresden, Dresden, Germany \vspace{-0.3cm}
\item[$^6$] These authors contributed equally and are joint first authors. \vspace{-0.3cm}
\item[$^7$] Corresponding author: fakhri@mit.edu
\end{itemize}

\vspace{1cm}
\normalsize
\renewcommand\contentsname{} 
\tableofcontents

\newpage
\section{Experimental methods}
\subsection{Preparation of starfish embryos}
Starfish (\textit{Patiria miniata}) were purchased from South Coast Bio-Marine LLC, Marinus Scientific, and Monterey Abalone Company. We kept the starfish in a salt water fish tank at 15$^{\circ}$C with appropriate water filtering systems, and fed them with shrimp every two weeks. Prior to fertilization, we extracted oocytes and sperm separately from female and male starfish, respectively, by first extracting the gonads via  small incisions made on the underside of the starfish near the stomach. The spermatophores were stored in Eppendorf tubes at 4$^{\circ}$C before use. We segmented the ovaries with scissors to release the oocytes. We then washed the oocytes twice with calcium-free sea water to prevent spontaneous maturation, and spread them into monolayers across multiple VWR non-treated sterilized 6-well culture plates (Supplier Part Number 10861-554) filled with filtered sea water (FSW). 
\\

To fertilize the embryos, we first added 10mM 1-methyladenine (1-MA) at a 1:1000 ratio to the oocyte culture to mature the oocytes. Typically, most starfish oocytes go through nuclear envelope breakdown within 2 hours of 1-MA addition, which we inspected using dissection scopes (Nikon SMZ745T). Within 2 hours of adding 1-MA, we added sperm to the oocyte culture for fertilization. The sperm culture was diluted such that the oocyte-to-sperm ratio was roughly 1:10. The fertilized starfish embryos were kept at 15$^{\circ}$C for the first 24 hours post-fertilization. Subsequently, we moved the starfish embryos to room temperature (approx. 20$^{\circ}$C) for experiments.

\subsection{Imaging starfish embryo dynamics}
\subsubsection{Imaging starfish embryos from the top}
\label{sssec:SI_imagingdyn}

Around 27 hours post-fertilization (hpf), we manually pipetted the swimming starfish embryos into VWR non-treated sterilized 24-well culture plates (Supplier Part Number 10861-558). For each well, about half of the cultured water was replaced with fresh FSW, with a total volume of around 2.5mL. The well plates were placed on a dissection scope (Nikon SMZ745T) at room temperature. Images were taken by a CMOS digital camera (Amscope MU500) attached to a fixed microscope adapter (AmScope FMA050) and then to the dissection scope eyepiece. We imaged the starfish embryo crystals for multiple hours at 1-10 seconds per frame. 
\\

\begin{figure}[h!]
  \centering
\includegraphics[width=\linewidth]{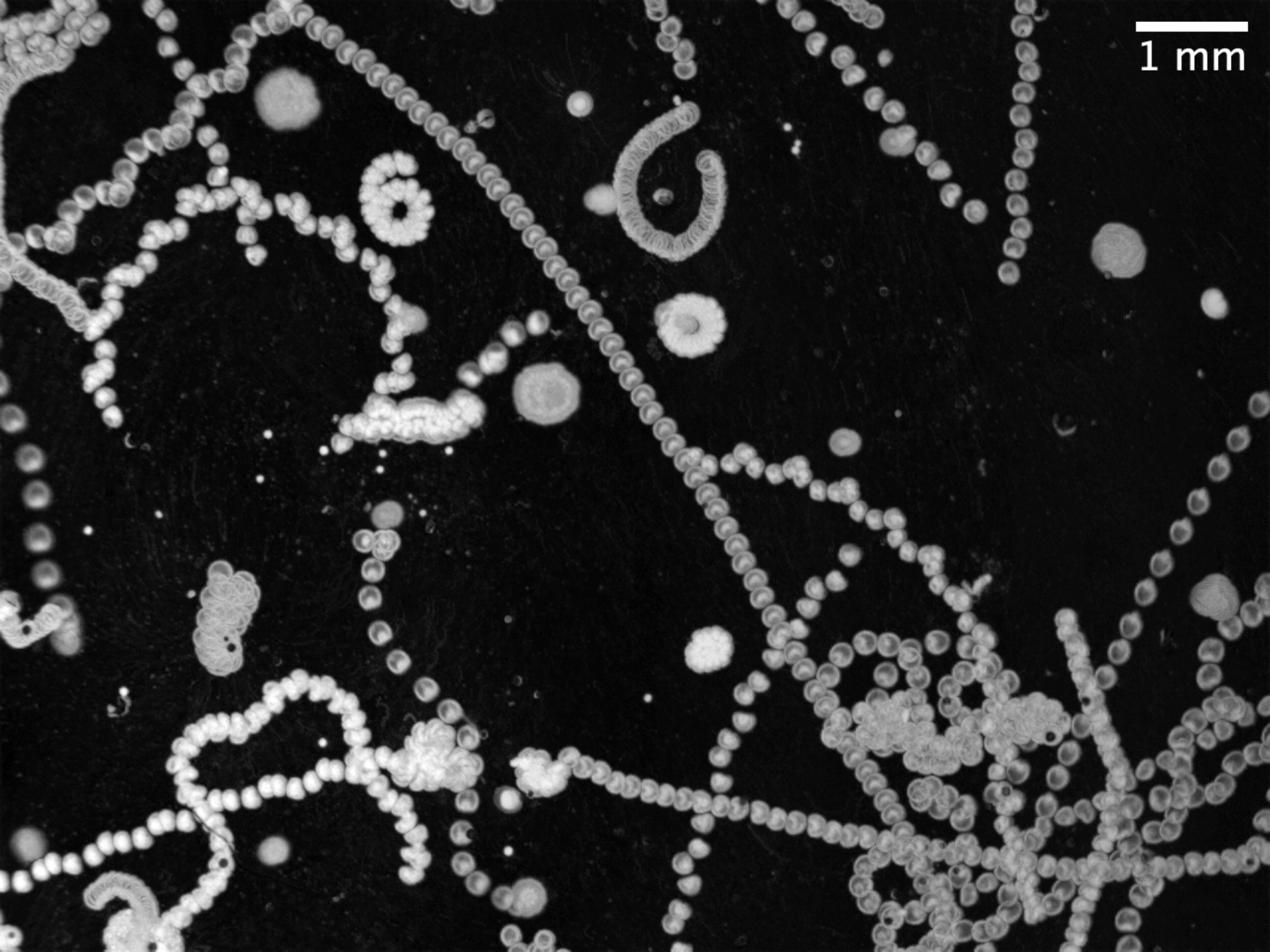}
    \caption{\textbf{ Helical and rectilinear swimming in starfish embryos.} Projections of swimming starfish embryos in the imaging plane over time reveal that embryos swim either in straight line (rectilinear) or corkscrew (helical) trajectories.}
  \label{figSI:corkscrew}
\end{figure}

To image swimming starfish embryos (see fig. \ref{figSI:corkscrew}), we transferred the embryos into Petri dishes and adjusted their concentration to ensure that we could observe an adequate number of embryos in the field of view. The total water level height was around 10 mm. Images were taken at 2 frames per second for multiple hours.

\subsubsection{Imaging starfish embryos from the side}
To obtain a side view of starfish embryos, we pipetted swimming embryos into a 25 ml tissue culture flask. We then held the flask in place between the objective and light source (fig. \ref{figSI:ExptSetup}a) of a dissection scope laid on its side. The images were taken at 2 frames per second for several hours.
\\

\subsection{Mechanical perturbation of living chiral crystals}
\label{SI:compressexpt}
\subsubsection{Uniaxial step compression experiments}
All mechanical compression experiments were performed after living crystal formation (about 45 hours hpf) and before disassembly of crystals (about 60 hpf). During compression experiments, the embryos were imaged under the dissection scope, with videos recorded at 1 frame per second. 
\\

After starfish embryos formed a living chiral crystal (LCC) that spanned one well of a 24-well plate, we carefully inserted four plastic walls to confine the crystal. Three of the walls were taped together, and the fourth wall (the ``piston'') was attached to a motor. All walls extended approximately 6 mm above and below the air-water interface. Following insertion of the walls, we let the crystal stabilize in the fluctuating state for at least 20 minutes before performing further perturbations. 
\\

\begin{figure}[h!]
  \centering
\includegraphics[width=\linewidth]{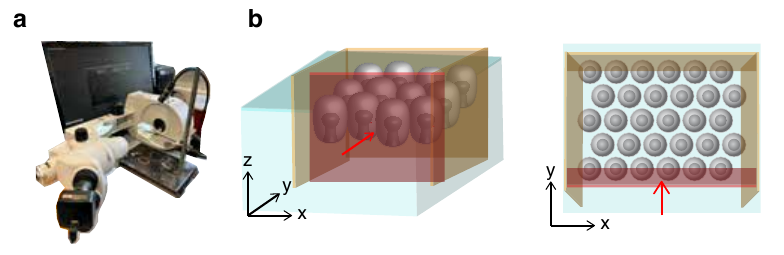}
    \caption{\textbf{Experimental setup and schematics.} \textbf{(a)} Side imaging setup. The dissection scope is laid on its side, and the sample is positioned between the light source and the camera. The computer screen in the back displays instantaneous images of the embryos captured from the side. \textbf{(b)} Schematic of the compression experiment setup. The embryos are confined between three static walls (yellow). Step compression is applied by a moving wall (red) attached to a motor. Left: side view; right: top view.}
  \label{figSI:ExptSetup}
\end{figure}

Mechanical step compressions were performed in a stepwise, uniaxial manner. Fast compressions were applied by motor-drive manipulator (Narishige MM-94), and slow compressions were applied by syringe pump (New Era Syringe Pump NE-300). The piston velocity was kept constant during each step compression. The total piston displacement of each step compression was between 1 to 3 mm for all compression experiments, corresponding to compression rates ranging between $\sim 10^{-4}$ to $\sim 0.03$ 1/s. We continuously recorded the crystal dynamics during and after each step compression. For serial step compressions, we ensured that the crystal was in the fluctuating state for at least 20 minutes before the next step compression (fig. \ref{figSI:ExptSetup}b).
\\

\section{Theory} \label{sec2}
\subsection{Motivation and explanation of equations in the continuum model}\label{sec2:1}
To explain the observed bistability and oscillatory dynamics in LCCs, we developed a phenomenological continuum theoretical model that incorporates odd elasticity as well as embryo tilt and precession dynamics. Our model draws upon the fact that while the LCC is effectively a two-dimensional solid, the embryos composing it are three-dimensional objects, whose orientation in space is described by an azimuthal angle $\phi$ and polar, or tilt angle $\theta$. Our model consists of the following equations, with repeated indices being summed over:

\begin{gather}\label{eq:S1}
    \frac{\partial u_x}{\partial t} = v_{0}  \sin\theta\cos\phi + C_{ixmn}\partial_{i}\partial_{m}u_n \\
\label{eq:S2}
    \frac{\partial u_y}{\partial t} = v_{0}\sin\theta\sin\phi + C_{iymn}\partial_{i}\partial_{m}u_n \\
\label{eq:S3}
    \frac{\partial \phi}{\partial t} = \frac{\Lambda_{AP}\sin(k\theta)\omega_s}{\left[\Lambda_{\perp}\sin\theta\cos(k\theta) - \Lambda_{AP}\cos\theta\sin(k\theta)\right]} \left(  1+\frac{6Rc}{d_0}\partial_k u_k \right) \\
\label{eq:S4}
    \frac{\partial \theta}{\partial t} = \gamma\theta \left(\frac{\theta}{\theta_c}-1\right)\left(1-\frac{\theta}{\theta_0}\right) + D\nabla^{2}\theta + \beta\partial_k u_k 
\end{gather}

In eqns. \ref{eq:S1} and \ref{eq:S2}, $u_j$ are components of the coarse-grained embryo displacement field $\mathbf{u}(\mathbf{r},t)$ as a function of position vector $\mathbf{r}$ and time $t$, and $C_{ijmn}$ is the elastic modulus tensor containing conventional as well as odd elastic moduli, normalized by the friction coefficient associated with an overdamped system. The coarse-grained orientational dynamics of embryos are described by the fields $\theta(\mathbf{r},t)$ and $\phi(\mathbf{r},t)$, corresponding to the embryos' tilt angle and azimuthal orientation, respectively. In the fluctuating state with $\theta = 0$, eqns. \ref{eq:S1} and \ref{eq:S2} reduce to the overdamped equations for a linear elastic solid $\partial u_j/\partial t = C_{ijmn}\partial_i\partial_m u_n$. In practice, coarse-graining over microscopic interactions also results in a transverse pre-stress that leads to global rotation of the living chiral crystal \cite{tan2022odd}. In the present work, we assume that all calculations are performed in the corotating frame, so that the pre-stress enters only via its contribution to odd elastic moduli. 
\\

The combined effect of finite tilt and precession is to endow embryos with an orientation-dependent self-propulsion, parametrized by the constant $v_0$. This self-propulsion speed captures the geometric contribution due to finite body length of embryos, as well as the active contribution due to the broken symmetry of the Stokeslet flows generated by tilted embryos. Assuming that the active torque generated by spinning embryos is balanced by the viscous torque exerted by the surrounding fluid, the coarse-grained precession frequency $\partial \phi / \partial t$ is given by eqn. \ref{eq:S3}, where $\Lambda_{AP}$ and $\Lambda_{\perp}$ are rotational drag coefficients about the AP axis and any of the two symmetry axes perpendicular to the AP axis, respectively, and $k$ is a phenomenological constant. We justify the choice of this functional form in section \ref{sec2:2}. In eqn. \ref{eq:S3}, $R$ is the embryo radius, $d_0$ is the surface-to-surface distance between embryos in an ordered LCC, and $c$ can be derived by coarse-graining the torque exchange interactions between embryos, as shown in section \ref{sec2:3}. Finally, eqn. \ref{eq:S4} is a phenomenological equation that describes local relaxation of the tilt angle field to stable values at $\theta = 0$ and $\theta = \theta_0 > 0$, which yields bistability, $\theta_c$ is an unstable tilt angle that separates the two stable angles, and $D$ and $\beta$ correspond to the strengths of tilt-tilt and tilt-displacement coupling, respectively.

\subsection{Relation between spin and precession frequencies}
\label{sec2:2}
Embryos bound near the air-water interface can generate active forces and torques via motion of their cilia. The balance between active forces and viscous drag determines the embryos' propulsion speed as well as rotation frequency. In the stable upright state, the active force $\mathbf{F_a}$ and active torque $\mathbf{\tau_a}$ generated by an embryo are both aligned along the lab frame $z$-axis, $\hat{z}$. Furthermore, the angular velocity vector of the spinning embryo, $\mathbf{\Omega} \parallel \mathbf{\tau_a}$. However, we also have a second stable state in which the embryo precesses about the lab frame $z$-axis with precession frequency $\omega_p$, in addition to spinning about its own AP axis with intrinsic rotation frequency $\omega_r$. This type of motion is possible if the active force and torque become misaligned. The simplest situation that can give rise to such embryo dynamics is the case in which $\mathbf{F_a}$,$\mathbf{\tau_a}$, and $\mathbf{\Omega}$ are no longer parallel, but always lie within the same plane, with $\mathbf{F_a}$ aligned with the AP axis of the embryo. 
\\

Under these assumptions, we can express the angular frequency in a reference frame fixed in space, but instantaneously aligned with the embryo's body frame, using the Euler angles $\theta$, $\phi$, and $\psi$ (See \cite{weisstein2009euler} for definitions and visualizations of the Euler angles). For the purposes of our problem, $\theta$ is the angle of the embryo's AP axis relative to the lab frame $z$-axis, the latter of which is normal to the air-water interface. Let $\hat{N}$ denote a unit vector along the line of intersection between the planes normal to the $z$-axis and the AP axis. The component of rotation frequency about $\hat{N}$ is the nutation frequency $\dot{\theta}$. $\phi$ is the angle between the lab frame $x$-axis and $\hat{N}$, and describes precession of the embryo about the $z$-axis, so the precession frequency $\omega_p = \dot{\phi}$. Finally, the angle $\psi$ describes the intrinsic rotation of the embryo about its AP axis; thus an embryo has intrinsic rotation frequency $\omega_r = \dot{\psi}$. These three Euler angles define a reference frame whose orthogonal axes $X$, $Y$, and $Z$ are oriented such that $Z$ is parallel to the AP axis. In what follows, we will assume that we can neglect the nutation frequency $\dot{\theta}$, and the AP axis therefore makes a constant angle $\theta_0$ with the lab frame $z$-axis. With this simplification, the components of angular velocity $\mathbf{\Omega}$ in the $XYZ$ reference frame is given by

\begin{gather}\label{eq:S5}
    \Omega_X = \omega_p \sin \theta_0 \sin \psi \\\label{eq:S6}
    \Omega_Y = \omega_p \sin \theta_0 \cos \psi \\\label{eq:S7}
    \Omega_Z = \omega_p \cos \theta_0 + \omega_r
\end{gather}

Since the torque vector $\mathbf{\tau_a}$ is coplanar with $\Omega$ and $\mathbf{F_a}$, and $\mathbf{F_a} \parallel \hat{Z}$, we have \\ $\mathbf{\tau_a} = \tau_0[\sin\theta_M\sin \psi,\sin\theta_M\cos \psi,\cos\theta_M]^{\mathrm{T}}$, where $\tau_0$ is the magnitude of the active torque, and $\theta_M$ is the angle specifying the mismatch between $\mathbf{F_a}$ and $\mathbf{\tau_a}$. Assuming that the active torque is balanced by rotational drag, we get the following matrix equation, assuming the embryo to be a prolate object: 
\begin{equation} \label{eq:S8}
    \begin{bmatrix}
        \Lambda_{\perp} & 0 & 0\\
        0 & \Lambda_{\perp} & 0\\
        0 & 0 & \Lambda_{AP}
    \end{bmatrix}
    \begin{bmatrix}
        \omega_p \sin \theta_0 \sin \psi \\
        \omega_p \sin \theta_0 \cos \psi \\
        \omega_p \cos \theta_0 + \omega_r
    \end{bmatrix}
    = 
    \begin{bmatrix}
        \tau_0 \sin \theta_M \sin \psi \\
        \tau_0 \sin \theta_M \cos \psi \\
        \tau_0 \cos \theta_M
    \end{bmatrix}
\end{equation}

By equating components, we arrive at the following expression for precession frequency $\omega_p$ as a function of intrinsic rotation frequency $\omega_r$

\begin{equation}\label{eq:S9}
    \omega_p = \frac{\omega_r \Lambda_{AP}\sin\theta_M}{\left(\Lambda_{\perp}\sin\theta_0\cos\theta_M - \Lambda_{AP}\cos\theta_0\sin\theta_M\right)}
\end{equation}

Extending the result obtained in eqn. \ref{eq:S9} to many-body systems is extremely challenging, as the microscopic mechanisms and dynamics associated with misalignment between active forces and torques are not experimentally extractable. Furthermore, $\theta_M$ will in general strongly influence the dynamics of the tilt angle $\theta$ via hydrodynamics as well as other near field effects that are difficult to capture. However, we note that most of the essential physics presented in this work hinges on the fact that there exists a finite stable tilt angle $\theta_0$, and is not particularly sensitive to the details of tilt dynamics. In our continuum model, therefore, we simply utilize the fact that $\theta_M\to 0$ as $\theta_0 \to 0$, and write $\theta_M = k\theta_0$, where $k$ is a constant. We further assume that the functional form given in eqn. \ref{eq:S9} holds for any nonzero $\theta$. The equation for the coarse-grained azimuthal angle field is then given by 

\begin{equation}\label{eq:S10}
    \frac{\partial\phi}{\partial t} = \frac{\omega_r \Lambda_{AP}\sin(k\theta)}{\left(\Lambda_{\perp}\sin\theta\cos(k\theta) - \Lambda_{AP}\cos\theta\sin(k\theta)\right)}
\end{equation}

The embryo spin frequency $\omega_r$ depends on the local strain due to hydrodynamic torque exchange interactions between embryos, as shown in the next section.

\subsection{Coarse-graining torque exchange interactions}\label{sec2:3}

In previous work, we showed that torque exchange interactions lead to the slowdown of embryo rotation \cite{tan2022odd}. Using results from lubrication theory \cite{kim2013microhydrodynamics,drescher2009dancing}, the variation of embryo rotation frequency with inter-embryo distance can be written as \cite{tan2022odd} 
\begin{gather} \nonumber
    \omega_{ri} = \omega_{r0} - \sum_{j \neq i} (\omega_{ri} + \omega_{rj})T_{\text{nf}}(|\mathbf{r}_i-\mathbf{r}_j|) \\ \label{eq:S11}
    \mathrm{where}\\ \nonumber
    T_{\text{nf}}(|\mathbf{r}_i-\mathbf{r}_j|) = \Bigg\{
    \begin{matrix}
        \tau_0\ln (d_c/d_{ij}) & (d_{ij} < d_c)\\
        0 & (d_{ij} \geq d_c)
    \end{matrix}
\end{gather}
where $\omega_{ri}$ is the rotation frequency of the $i^{\mathrm{th}}$ embryo in an LCC, $\omega_{r0}$ is the intrinsic rotation frequency of individual embryos bound at the air-water interface, $\tau_0$ is a dimensionless parameter quantifying the strength of torque exchange, $d_c$ is the range of torque exchange interactions, and $d_{ij} = |\mathbf{r}_i-\mathbf{r}_j|-2R$ is the surface to surface distance between embryos $i$ and $j$. 
\\

While the torque exchange interactions in eqn. \ref{eq:S11} have an intrinsic many-body character, we implement a simple coarse-graining procedure based on a local mean field approximation. We begin by assuming that the rotation frequency varies slowly in space, and make a local mean field approximation by replacing the frequency of neighboring particles in eqn. \ref{eq:S11} by the local average rotation frequency $\omega_r$. Further, using the functional form for the near-field torque $T_{\text{nf}}(|\mathbf{r}_i-\mathbf{r}_j|)$ given in equation \ref{eq:S11}, we can write 

\begin{equation} \label{eq:S12}
    \omega_r = \frac{\omega_{r0}}{1+2\tau_0\sum_{j \neq i}\text{ln}(d_c/d_{ij})}
\end{equation}
where $d_{ij}$ is the surface to surface distance between embryos $i$ and $j$. The sum only runs over nearest neighbors as the torque exchange term is only due to lubrication interactions, which are very short-ranged. Let $d_0$ be the equilibrium surface to surface distance between neighboring embryos in a perfectly ordered 2D triangular lattice, and $2R$ be the embryo diameter, such that the equilibrium lattice constant is $2R + d_0$. Assuming that $d_0$ as well as $d_{ij}$ are small compared to $2R$, using standard elasticity theory \cite{landau1986theory}, we get

\begin{equation} \label{eq:S13}
    \frac{d_{ij}}{d_0} = 1 + \frac{u_{ij}^{kl}}{2Rd_0}dx^kdx^l = 1 + \delta_{ij}
\end{equation}

where $u_{ij}^{kl} = (1/2)(\partial_l u_{ij}^{k} + \partial_k u_{ij}^{l})$ is the $kl$-th component of the local strain tensor associated with embryos $i$ and $j$, and $dx^m$ are components of the relative position vector between embryos $i$ and $j$ in the undeformed lattice. Further, the small deformation approximation implies that $\delta_{ij} << 1$. Using eqn. \ref{eq:S13} and approximating $\text{ln}(1+\delta_{ij}) \approx \delta_{ij}$, eqn. \ref{eq:S12} becomes

\begin{equation} \label{eq:S14}
    \omega_r = \frac{\omega_{r0}}{1+2\tau_0 N_c \text{ln}(d_c/d_0) - 2\tau_0\sum_{j \neq i} \delta_{ij}}
\end{equation}

where $N_c = 6$ is the coordination number of the 2D hexagonal lattice. 
Defining $\omega_s = \omega_{r0}/(1+2\tau_0 N_c \text{ln}(d_c/d_0))$, and $c = 2\tau_0/(1+2\tau_0 N_c \text{ln}(d_c/d_0))$, and using $c\sum_{j \neq i} \delta_{ij} \ll 1$, we can rewrite the previous equation as

\begin{equation} \label{eq:S15}
    \omega_r = \omega_s \left(1+c\sum_{j \neq i} \delta_{ij} \right)
\end{equation}

As the sum in eqn. \ref{eq:S15} runs only over nearest neighbors, we can use the lattice symmetries to explicitly write out the six terms in the sum. Since the same exercise can be repeated for any embryo, we drop the label $i$ for convenience:

\begin{equation} \label{eq:S16}
\begin{aligned}
   \delta_1 = \frac{2R}{d_0}u_1^{11};  \hspace*{1cm} \delta_4 = \frac{2R}{d_0}u_4^{11};\\
  \delta_2 = \frac{2R}{d_0}\left(\frac{1}{4}u_2^{11} + \frac{\sqrt{3}}{4}(u_2^{12} + u_2^{21}) + \frac{\sqrt{3}}{4}u_2^{22} \right);\\
   \delta_5 = \frac{2R}{d_0}\left(\frac{1}{4}u_5^{11} + \frac{\sqrt{3}}{4}(u_5^{12} + u_5^{21}) + \frac{\sqrt{3}}{4}u_5^{22} \right);\\
   \delta_3 = \frac{2R}{d_0}\left(\frac{1}{4}u_3^{11} - \frac{\sqrt{3}}{4}(u_3^{12} + u_3^{21}) + \frac{\sqrt{3}}{4}u_3^{22} \right);\\
   \delta_6 = (\frac{2R}{d_0}\left(\frac{1}{4}u_6^{11} - \frac{\sqrt{3}}{4}u_6^{12} + u_6^{21}) + \frac{\sqrt{3}}{4}u_6^{22} \right);
\end{aligned}
\end{equation}

In our simple coarse grained theory, we can replace the local strains associated with different neighbors by components of the average local strain tensor. Thus, $\sum_{j} \delta_{j} = (2R/d_0)(3u^{11}+3u^{22}) = (6R/d_0)\partial_k u_k$, where the index $k$ is summed over. We see that components of the shear strain get cancelled out, and we are left with the divergence of the strain tensor alone. Thus, the coarse grained rotation frequency field is given by 

\begin{equation}\label{eq:S17}
    \omega_r = \omega_s \left(1+\frac{6Rc}{d_0}\partial_k u_k \right)
\end{equation}

\subsection{Range expansion of tilted region}
\label{sec2:4}
The transition between the fluctuating state with $\theta=0$ and the oscillatory state with stable, finite tilt angle $\theta = \theta_0$ is accompanied by spatial tilt dynamics. In particular, the transition of the LCC system to the oscillatory state via expansion of the initial oscillating region in the main text (Fig. 1) can be understood as an invasion of the region in the fluctuating state by the region in the oscillatory state. Mathematically, this spatial takeover by the oscillatory state is equivalent to range expansion of tilt population $\theta(\mathbf{r},t)>0$ into the empty space with $\theta=0$ in the limit of negligible tilt-displacement coupling $\beta$. With $\beta \to 0$, equation \eqref{eq:S4} simplifies to:
\begin{equation} \label{eq:Sx1}
\begin{aligned}
    \frac{\partial \theta}{\partial t} = \gamma\theta \left(\frac{\theta}{\theta_c}-1\right)\left(1-\frac{\theta}{\theta_0}\right) + D\nabla^{2}\theta
\end{aligned}
\end{equation}
We can rewrite the equation as a growth and dispersal of the normalized tilt angle $p \equiv \theta / \theta_0$:
\begin{equation} \label{eq:Sx2}
\begin{aligned}
    \frac{\partial p}{\partial t} = D\nabla^{2}p + (r_0 + r_p p) p (1-p)
    \\ p = \theta/\theta_0, \quad r_0 = -\gamma, \quad r_p = \gamma \theta_0/\theta_c.
\end{aligned}
\end{equation}
In this model, while the base ``growth rate'' of the tilt population $r_0=-\gamma$ is negative when $p$ is small, the net growth rate $(r_0 + r_p)p$ increases as $p$ increases. This is the bistability we observe in tilt angle $\theta$, in which a sufficiently large tilt stabilizes at nonzero value $\theta = \theta_0$.

Following Li and Petrovskii~\cite{petrovskii2005exactly}, we find that the large tilt region ($\theta = \theta_0$) may be able to expand if the ratio between stable and unstable tilt angles satisfies
\begin{equation} \label{eq:Sx3}
\begin{aligned}
    \frac{\theta_c}{\theta_0} < \frac{1}{2}
\end{aligned}
\end{equation}

Once the above inequality is satisfied, similar to the bistability in local tilt angle, the region of nonzero tilt can expand if its initial size is sufficiently larger than the characteristic length scale $\sqrt{D/\gamma}$. The initial distribution of tilt angle $p_c(x)$ that determines the critical size of the tilt region for spatial expansion satisfies
\begin{equation} \label{eq:Sx4}
\begin{aligned}
  D\nabla^{2}p_c + (r_0 + r_p p_c) p_c (1-p_c)=0
\end{aligned}
\end{equation}
We linearize eqn. \ref{eq:Sx4} near $p_c=1$ by writing $p_c(x) = 1 - \epsilon e^{-x/L_0}$, where $L_0$ represents the initial size of the large tilt region. From equation \eqref{eq:Sx4}, we get
\begin{equation} \label{eq:Sx5}
\begin{aligned}
\frac{\theta_c}{\theta_0} = \frac{1}{1+ \frac{D}{\gamma L_0^2}}
\end{aligned}
\end{equation}
In the end, the tilt ratio $\frac{\theta_c}{\theta_0} $ should satisfy two criteria in order to enable the expansion of the oscillating region:
\begin{equation} \label{eq:Sx6}
\begin{aligned}
\frac{\theta_c}{\theta_0} < \min\left(\frac{1}{2},\ \frac{1}{1+ \frac{D}{\gamma L_0^2}}\right)
\end{aligned}
\end{equation}
fig. \ref{fig:oddwave_PD} - \ref{fig:beta1_PDs} show that no tilt oscillations are permitted past this boundary, indicating agreement with the above result.

\subsection{Dispersion relation and prediction of optical bands}
\label{sec2:5}
To understand how tilt and precession dynamics influence vibrational modes of the LCC, we calculated the dispersion relation analytically in the limit of weak torque exchange ($c \to 0$), and weak tilt-displacement coupling ($\beta \to 0$). In this limit, tilt and displacement dynamics decouple, the tilt angle evolves to a constant value $\theta = \theta_0$ uniformly in space, and eqns. \ref{eq:S1}-\ref{eq:S3} reduce to the following simplified set of equations:

\begin{gather}\label{eq:S18}
    \frac{\partial u_x}{\partial t} = v_{m} \cos\phi + C_{ixmn}\partial_{i}\partial_{m}u_n \\
    \label{eq:S19}
    \frac{\partial u_y}{\partial t} = v_{m} \sin\phi + C_{iymn}\partial_{i}\partial_{m}u_n \\
    \label{eq:S20}
    \frac{\partial \phi}{\partial t} = \omega_p
\end{gather}

where $v_m = v_0\sin\theta_0$, and $\omega_p = \omega_s[\Lambda_{AP}\sin k\theta_0]/[\Lambda_{\perp}\sin\theta_0\cos(k\theta_0) - \Lambda_{AP}\cos\theta_0\sin(k\theta_0)]$ is the precession frequency. 

Differentiating eqns. \ref{eq:S18}-\ref{eq:S19} once with respect to time, we get

\begin{gather}\label{eq:S21}
    \frac{\partial^2 u_x}{\partial t^2} = -v_{m} \sin\phi\frac{\partial \phi}{\partial t} + C_{ixmn}\partial_{i}\partial_{m}\partial_t u_n \\
    \label{eq:S22}
    \frac{\partial^2 u_y}{\partial t^2} = v_{m} \cos\phi\frac{\partial \phi}{\partial t} + C_{iymn}\partial_{i}\partial_{m}\partial_t u_n 
\end{gather}

We can eliminate $\phi$ by substituting eqn. \ref{eq:S20} in eqns. \ref{eq:S21} and \ref{eq:S22}, multiplying eqns. \ref{eq:S18} and \ref{eq:S19} by $\omega_p$ and considering appropriate linear combinations of the resulting equations. We then arrive at the following equation for displacements:

\begin{equation}\label{eq:S23}
    \frac{\partial^2 u_j}{\partial t^2} + \omega_p\epsilon_{jk}\frac{\partial u_k}{\partial t} = C_{ijmn}\partial_i\partial_m\partial_t u_n + \omega_p\epsilon_{jk}C_{ikmn}\partial_i\partial_m u_n
\end{equation}

where $\epsilon_{jk}$ is the Levi-Civita symbol in 2D.  Rewriting the above equations in terms of Fourier transforms 
$$u_j (\mathbf{r},t) = \int \Tilde u_j(\mathbf{q},\omega) e^{i(\mathbf{q} \cdot \mathbf{r} - \omega t)} d\mathbf{q} d\omega$$
with $j\in\{x,y\}$, and taking as our elastic moduli tensor \cite{scheibner2020odd}
\begin{align}
    \begin{split}
        C_{ijmn} &= B\delta_{ij}\delta_{mn} + \mu(\delta_{in} \delta_{jm} + \delta_{im} \delta_{jn} - \delta_{ij} \delta_{mn}) \\
        &+ \frac 12 K_0 (\epsilon_{im} \delta_{jn} + \epsilon_{in} \delta_{jm} + \epsilon_{jm} \delta_{in} + \epsilon_{jn} \delta_{im}) - A \epsilon_{ij} \delta_{mn},
    \end{split}
\end{align}
we get acoustic ($\omega_{\mathrm{ac}}$) as well as optical ($\omega_{\mathrm{op}}$) branches of the dispersion relation, given by

\begin{gather} \label{eq:S24}
    \omega_{\mathrm{op}} = \pm \omega_p \\ 
    \label{eq:S25}
    \omega_{\mathrm{ac}} = -i\left[\ \frac{B}{2} + \mu \pm \sqrt{\left(\frac{B}{2}\right)^2-K_0A-K_0^2} \right]q^2
\end{gather}

The acoustic modes are identical to the modes of an overdamped linear odd elastic solid \cite{scheibner2020odd}. The main effect of tilt and precession is thus to introduce optical vibrational modes, the dispersion of which is independent of the wave vector $\mathbf{q}$ in the long wavelength limit. 

\subsubsection{Note on the existence of the optical band}
\label{SI:opticalband}
Even though the dispersion relation predicts the existence of an optical band, it is not guaranteed that such a band will appear unambiguously within the data. As our system is not in thermal equilibrium, which modes are excited, and hence detected, within our data depends on the initial conditions. To see this, it is instructive to begin again from eqns. \ref{eq:S18}-\ref{eq:S20}.  Integrating eqn. \ref{eq:S20} yields the solution for the azimuthal angle $\phi(\mathbf{r},t) = \phi_0(\mathbf{r}) + \omega_p t$; substituting this expression into eqns. \ref{eq:S18} and \ref{eq:S19} gives
\begin{align}
    \begin{split}
        \frac{\partial u_x}{\partial t} &= C_{ixmn}\partial_{i}\partial_{m}u_n + f(\mathbf{r},t)\\
    \frac{\partial u_y}{\partial t} &= C_{iymn}\partial_{i}\partial_{m}u_n + g(\mathbf{r},t) \\
    \end{split}
\end{align}

where
\begin{align}
    \begin{split}
        f(\mathbf{r},t) &\equiv v_0 \sin \theta_0 \cos \phi = v_m(\cos \phi_0 \cos \omega_pt - \sin \phi_0 \sin \omega_pt) \\
        g(\mathbf{r},t) &\equiv v_0 \sin \theta_0 \sin \phi = v_m(\cos \phi_0 \sin \omega_pt + \sin \phi_0 \cos \omega_pt)
    \end{split}
\end{align}
can be interpreted as temporal forcing variables. After taking spatial Fourier transforms, the above equations take the form of inhomogeneous ODEs, the solution to which can be obtained by first solving the homogeneous set of equations
\begin{align}
    \begin{split}
        \frac{\partial u_x}{\partial t} &= C_{ixmn}\partial_{i}\partial_{m}u_n \\
    \frac{\partial u_y}{\partial t} &= C_{iymn}\partial_{i}\partial_{m}u_n\\
    \end{split}
\end{align}
These equations describe purely elastic dynamics and correspond to the dispersion relation of an odd elastic solid \cite{scheibner2020odd}, as given in eqn. \ref{eq:S25}. This immediately tells us that the forcing terms $f(\mathbf{r},t)$ and $g(\mathbf{r},t$) must be nonzero in order to obtain any result beyond odd elasticity in our model, and in particular that the Fourier transforms $\Tilde{f}$ and $\Tilde{g}$ must be nonzero at a given $\mathbf{q}$ in order for any precession-related dynamics to be present: 
\begin{align}
    \begin{split}
        \Tilde{f}(\mathbf{q},\omega) &= \pi v_0 (\Tilde{\phi}_{0,+} \delta_{\omega_{p}, \omega} + \Tilde{\phi}_{0,-} \delta_{\omega_p,-\omega})  \\
        \Tilde{g}(\mathbf{q}, \omega) &= -i\pi v_0 (\Tilde{\phi}_{0,+} \delta_{\omega_{p},\omega} - \Tilde{\phi}_{0,-} \delta_{\omega_p,-\omega})
    \end{split}
\end{align}
where $\Tilde{\phi}_{0,\pm} = \int e^{\pm i\phi_0(\mathbf{r})} e^{-i\mathbf{q} \cdot \mathbf{r}} d\mathbf{r}$. Note that the amplitudes of these Fourier transforms depend on the initial configuration of $\phi(\mathbf{r}, t=0) = \phi_0(\mathbf{r})$. Thus the optical branch only appears at a given $\mathbf{q}$ if the Fourier coefficient of $\phi_0$ is nonzero at that wavevector, and the substitution resulting in eqn. \ref{eq:S23} implicitly assumes that the Fourier mode of $\phi_0$ corresponding to $\mathbf{q}$ exists, though this is not guaranteed. As such, the existence of a full optical band in an LCC depends on the initial conditions of the azimuthal orientation field $\phi(\mathbf{r},t)$.

\subsection{Selective excitation of oscillations at intermediate compression rates}
\label{sec2:6}
In our step compression experiments, we observe that starting from the non-oscillatory state, oscillations can be selectively excited over a narrow range of applied compression rates. The compression step often leads to a disruption of the crystal and introduces strong nonlinearities including plastic flow as well as escape of embryos from the air-water interface into the bulk fluid, which precludes analytical treatment of the imposed strain field. Instead, we make theoretical progress by assuming that the main effect of compression is to set the initial conditions for the evolution of the system's displacement field at the end of the compression step. 

Since our goal is to understand how oscillations can be excited from the non-oscillatory state, our strategy is to examine the dynamics of our system by linearizing the system of equations described in eqns. \ref{eq:S1}-\ref{eq:S4} near $\theta = 0$.  

In the $\theta \to 0$ limit, $ \sin{\theta} \to 0$, and we can therefore ignore the self-propulsion term in eqns. \ref{eq:S1}-\ref{eq:S2} in comparison to the elastic terms. Under this assumption,  eqn. \ref{eq:S3} does not contribute to displacement dynamics. Furthermore, we linearize the reaction term in eqn. \ref{eq:S4}, which yields the following linearized system of equations:

\begin{gather} \label{eq:S26}
    \frac{\partial u_x}{\partial t} = C_{ixmn}\partial_{i}\partial_{m}u_n \\ \label{eq:S27}
    \frac{\partial u_y}{\partial t} = C_{iymn}\partial_{i}\partial_{m}u_n \\ \label{eq:S28}
     \frac{\partial \theta}{\partial t} = -\gamma\theta + D\nabla^{2}\theta+ \beta\partial_k u_k
\end{gather}

Fourier transforming the above equations in space, we get 

\begin{gather} \label{eq:S29}
    \frac{\partial \Tilde{u}_x}{\partial t} = C_{ixmn}q_{i}q_{m}\Tilde{u}_n \\ \label{eq:S30}
    \frac{\partial \Tilde{u}_y}{\partial t} = C_{iymn}q_{i}q_{m}\Tilde{u}_n \\ \label{eq:S31}
     \frac{\partial \Tilde{\theta}}{\partial t} = -\gamma\Tilde{\theta} + Dq^2\Tilde{\theta} + i\beta q_k \Tilde{u}_k
\end{gather}

where $q_i$, $q_m$ and $q_k$ denote components of the wave vector $\mathbf{q}$. eqns. \ref{eq:S29}-\ref{eq:S31} constitute a homogeneous linear system of differential equations of the form 

\begin{equation} \label{eq:S32}
    \frac{\partial \Tilde{\mathbf{x}}}{\partial t} = \mathbb M \Tilde{\mathbf{x}}
\end{equation}

where $\Tilde{\mathbf{x}} = [\Tilde{u}_x \hspace{0.1cm} \Tilde{u}_y \hspace{0.1cm} \Tilde{\theta}]^\text{T}$, $[\hspace{0.1cm}]^{\text{T}}$ denoting the matrix transpose. This system of equations can be solved exactly by finding the eigenvalues $\lambda_i(\mathbf{q})$ and eigenvectors $\mathbf{v}_i(\mathbf{q})$ of the matrix $\mathbb M$. The general solution to eqn. \ref{eq:S32} is of the form 

\begin{equation} \label{eq:S33}
    \Tilde{\mathbf{x}} = \sum_{i=1}^{3} C_i(\mathbf{q})\mathbf{v}_i(\mathbf{q}) e^{\lambda_i(\mathbf{q})t}
\end{equation}

where the functions $C_i(\mathbf{q})$ have to be determined from the initial conditions $\Tilde{\mathbf{x}}(t=0) = [\Tilde{u}_{x0} \hspace{0.1cm} \Tilde{u}_{y0} \hspace{0.1cm} \Tilde{\theta}_0]^\text{T}$. 

For the system described by eqns. \ref{eq:S29}-\ref{eq:S31}, the eigenvalues are given by 

\begin{gather}\label{eq:S34}
        \lambda_1 = -(\gamma + Dq^2)\\\label{eq:S35}
        \lambda_2 = -q^2\left(B+2\mu+\sqrt{B^2-4K_0^2-4AK_0}\right)\\\label{eq:S36}
        \lambda_3 = -q^2\left(B+2\mu-\sqrt{B^2-4K_0^2-4AK_0}\right)
\end{gather}

where $B$ and $\mu$ are the bulk and shear modulus, respsectively, $A$ is the odd elastic modulus that couples compressions to rotations, and $K_0$ is the odd elastic modulus that couples the two independent shear modes. While $\lambda_2$ and $\lambda_3$ describe the relaxation of the displacement field of a linear odd elastic solid \cite{scheibner2020odd}, $\lambda_1$ describes the relaxation of the tilt angle field.

\subsubsection{Plane wave displacement initial conditions}
Before discussing the effect of compression on the initial displacement field, it is instructive to solve eqn. \ref{eq:S33} for the simple case of a plane wave along the x-direction, such that $u_x(x,y) = \sin(q_0x)$ and $u_y(x,y) = \cos(q_0x)$. Fourier Transforming in space, we get $\Tilde{u}_{x0} = -2\pi^2 i\delta(q_y)(\delta(q_x+q_0)-\delta(q_x-q_0))$, and $\Tilde{u}_{y0} = 2\pi^2 \delta(q_y)(\delta(q_x+q_0)+\delta(q_x-q_0))$. Furthermore, since we are interested in the conditions under which oscillations are excited from the non-oscillatory state, we choose the initial tilt angles to be zero everywhere, i.e. $\Tilde{\theta}_0 = 0$. Plugging these initial conditions in eqn. \ref{eq:S33} to determine $C_i(\mathbf{q})$, and taking the inverse Fourier Transform in space, we can find the solutions for $u_x(x,y,t)$, $u_y(x,y,t)$, and $\theta(x,y,t)$ in space and time. From the solution for $\theta(x,y,t)$ in particular, we can figure out whether oscillations can be excited at specific spatial locations in the system. To see this, consider the solution for $\theta(x,y,t)$ at the origin, i.e. $(x=0,y=0)$, for the initial conditions given above. The exact solution in this case is

\begin{gather} \label{eq:S37}
   \theta(0,0,t;q_0) = \frac{\beta q_0}{\alpha_1 q_0^4 + \alpha_2 q_0^2 + \gamma^2} \hspace{0.5cm} \times \\ \nonumber \bigg[e^{\frac{-q_0^2}{2}(B+2\mu)t}\left( (\gamma+(D-\mu)q_0^2)\text{cosh}\left(\frac{\Delta q_0^2 t}{2}\right) - \left(\frac{B\gamma + \alpha_3 q_0^2}{\Delta}\right)\text{sinh}\left(\frac{\Delta q_0^2 t}{2}\right)\right) \\ \nonumber 
   - (\gamma+(D-\mu)q_0^2)e^{-(\gamma+Dq_0^2)t} \bigg] \nonumber
\end{gather}

where
\begin{gather*}
   \alpha_1 = D^2 - 2D\mu - BD + K_0^2 + AK_0 + \mu^2 + B\mu \\
   \alpha_2 = \gamma(2D - 2\mu - B) \\
   \alpha_3 = BD - 2K_0^2 -2AK_0 - B\mu \\
   \text{and} \hspace{0.2cm} \Delta = \sqrt{B^2 - 4K_0^2 - 4AK_0} 
\end{gather*} 

We have included $q_0$ in the argument of $\theta$ to indicate that the solution in eqn. \ref{eq:S37} follows from an initial condition corresponding to a plane wave of wavelength $2\pi/q_0$. We can easily check that $\theta(0,0,0;q_0) = 0$, as our initial condition demands. In our simplification of the system we ignored the nonlinear growth term of eqn. \ref{eq:S4}, and thus the tilt angle has no choice but to eventually relax to the fluctuating state, that is, $\theta \to 0$ as $t \to \infty$. 

Crucially, however, $\theta$ can transiently become positive, and this allows us to determine the conditions under which oscillations can be excited locally starting from the non-oscillatory state. This condition is given by $\text{max}(\theta) > \theta_c$. Physically, this condition implies that oscillations can be excited whenever the coupling between displacements and tilt is strong enough to push $\theta$ beyond the unstable fixed point at $\theta = \theta_c$ (eqn. \ref{eq:S4}) and into the domain of attraction of the fixed point at $\theta = \theta_0$. 

Eqn. \ref{eq:S37} also has a non-monotonic dependence on $q_0$. We can easily see that at any finite time, $\theta \to 0$ as $q_0 \to 0$ and $\theta \to 0$ as $q_0 \to \infty$. We as such expect that $\theta$ is maximized at some finite intermediate value of $q_0$. This implies that oscillations are most likely to be excited when the Fourier decomposition of the initial displacement profile contains modes of intermediate spatial frequency. At a fundamental level, this result follows from the fact that as $\theta \to 0$, the local relaxation term as well as diffusion term in eqn. \ref{eq:S31} favor the non-oscillatory state, whereas the tilt-displacement coupling term is the only one that can promote oscillations. Since local tilt relaxation dominates at small $q$, while diffusion dominates at large $q$, excitation of oscillations is most likely at intermediate $q$. This result is crucial to understand selective excitation of oscillations at intermediate compression rates, as we show in the following sections. 

\subsubsection{Dominant mode approximation}
As mentioned earlier, we assume that the applied compression rate determines the initial displacement field for our system. We can always express this initial displacement field as a superposition of plane waves with distinct wave vectors $\mathbf{q}$. Empirically, we observe that very low compression rates lead to the excitation of low energy, long wavelength (acoustic) elastic modes, whereas high compression rates lead to much more localized deformations, with a higher frequency of plastic neighbor exchange events (fig. \ref{fig:NNexchangeStrainrate}). These empirical observations suggest that the wave vector corresponding to the dominant spatial mode excited by the compression protocol is directly proportional to the compression rate $\Dot{\epsilon}$. By substituting $q_0 = \alpha\Dot{\epsilon}$ in eqn. \ref{eq:S37}, with $\alpha$ being a phenomenological proportionality constant, we can derive an exact expression for the range of compression rates over which excitation of sustained oscillations is possible. While this `dominant mode approximation' is a useful toy model, the key physical ingredients are retained for initial conditions with a more complex $q$ dependence, as we illustrate in the next section. 

\subsubsection{Gaussian kernel approximation}
\label{ssec:SI_gauskernel}
\begin{figure}[h!]
  \centering
\includegraphics[width=0.4\linewidth]{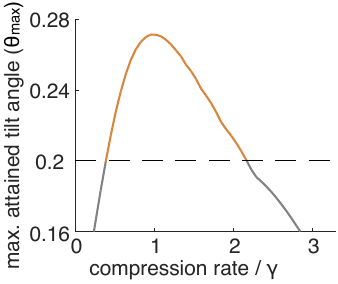}
    \caption{\textbf{Theoretical prediction of selective excitation of oscillations at intermediate compression rates.} Maximum achievable tilt angle $\theta_{max}$ for linearized dynamics as a function of the applied compression rate. Values of the compression rate (measured in simulation units set by $\gamma=1$) where $\theta_{max} > \theta_c$ (dashed line) indicate the regime where the system can sustain stable oscillations, given the dynamics described in eqns. \ref{eq:S26} - \ref{eq:S29}. See section 2.6.3 for details of parametrization.}
  \label{figSI:linear_compression_theory}
\end{figure}

Rather than a single dominant mode, it is more realistic to assume that a given compression step excites a band of spatial modes, whose mean increases linearly with the compression rate $\Dot{\epsilon}$. In this case, the initial displacement field can be defined as a superposition of plane waves with wave vectors $\mathbf{q_0}$ whose amplitudes are specified by a kernel $A(q_0)$, which depends on the compression rate. As a specific example, we define a Gaussian amplitude kernel $A(q_0)$ for each $q_0$ such that

\begin{equation} \label{eq:S38}
    A(q_0) = \text{exp}\left[\frac{-(q_0-\alpha\Dot{\epsilon})^2}{2\sigma^2}\right]
\end{equation}

where $\alpha$ and $\sigma$ are phenomenological parameters. Assuming such a kernel, the Fourier Transform of the initial displacement field now takes the form 

\begin{gather} \label{eq:S39}
    \Tilde{u}_{x0}(q_x,q_y) = -2\pi^2 i\int_{q_{\text{min}}}^{q_{\text{max}}}\text{exp}\left[\frac{-(q_0-\alpha\Dot{\epsilon})^2}{2\sigma^2}\right]\delta(q_y)(\delta(q_x+q_0)-\delta(q_x-q_0))dq_0  \\ 
    \Tilde{u}_{y0}(q_x,q_y) = 2\pi^2 \int_{q_{\text{min}}}^{q_{\text{max}}}\text{exp}\left[\frac{-(q_0-\alpha\Dot{\epsilon})^2}{2\sigma^2}\right]\delta(q_y)(\delta(q_x+q_0)+\delta(q_x-q_0))dq_0 \nonumber
\end{gather}

where $q_{\text{min}} = 2\pi/L$, $L$ being the system size, is the minimum possible value of $q_0$, and $q_{\text{max}} = \pi/(R)$, $R$ being the embryo radius, is the maximum possible value of $q_0$. Here, we have assumed that all wave vectors, regardless of their magnitude, point in the same direction ($x$-axis). Relaxing this assumption complicates the mathematics, and will in general influence the probability of exciting oscillations in a given location. However, the nonmonotonicity as a function of $q_0$ and time, which is the essential qualitative feature that we are interested in, remains unchanged. Likewise, to obtain the results we are interested in, we are not restricted to use Gaussian kernels, and several different functional forms for $A(q_0)$ will yield qualitatively similar results. Based on the initial conditions in eqn. \ref{eq:S39}, and once again assuming $\Tilde{\theta}_0 = 0$ uniformly in space, the solution for $\theta$ at the origin ($x=0,y=0$) takes the form

\begin{equation} \label{eq:S40}
    \theta(0,0,t) = \int_{q_{\text{min}}}^{q_{\text{max}}}\text{exp}\left[\frac{-(q_0-\alpha\Dot{\epsilon})^2}{2\sigma^2}\right] \theta(0,0,t;q_0) dq_0
\end{equation}

where $\theta(0,0,t;q_0)$ is given by eqn. \ref{eq:S37}. Performing the integration over $q_0$ can be quite tedious, and analytical solutions, whenever available, typically assume highly cumbersome forms. It is therefore more instructive to integrate eqn. \ref{eq:S40} numerically, to see that oscillations can indeed be locally excited only over a finite range of compression rates. As an illustrative example, we have performed this exercise with the following parameters: $\gamma = D = \beta = 1$, $B = 0.3$, $\mu = K_0 = A = B/2$, $R = 0.5$, $L = 100$, $\sigma = 0.1$, $\theta_c = 0.2$, and $\alpha = 1$, where the meanings of all parameters follow from eqns. \ref{eq:S1}-\ref{eq:S4}, \ref{eq:S34}-\ref{eq:S35}, \ref{eq:S38}, and \ref{eq:S39}. The results are plotted in fig. \ref{figSI:linear_compression_theory}, and we see that there is a well-defined range of compression rates over which oscillations can be excited at the origin. Thus, we can conclude that selective excitation of oscillations over a narrow range of compression rates arises due to the coupling between tilt and displacement fields as well as the manner in which the compression rate affects the spectrum of spatial modes in the displacement field.  

\section{Data analysis}

\subsection{Starfish embryo image segmentation and centroid tracking}
\label{sec:SegmentTrack}

To identify and segment individual starfish embryos within each top view image, we use \texttt{Cellpose} \cite{stringer2021cellpose}, a deep-learning-based segmentation algorithm, with the ``cyto2'' pre-trained model and appropriate embryo diameter for each dataset. To track embryo positions over time, we use \texttt{Trackpy} \cite{trackpy,crocker1996methods} and adjust the tracking parameters to optimize the number of particles with long tracks. We primarily adjust the search range and memory parameters, which control the distance a segmented object can travel within a single time step, and for how many frames an object can remain undetected while still being considered part of the same trajectory, respectively.
\\ 

To segment individual starfish embryos from side view images (Fig 1a), we first performed image-intensity thresholding with Otsu's method. We then identified the embryo boundaries using \texttt{skimage}. To simplify the pixelated boundaries, we drew the embryo outline based on the identified boundaries by hand.
\\

\subsection{Time series of averaged velocity and fraction of oscillating embryos} 
\label{SI:avgvelfrac}

This section describes how we obtained time series of the averaged velocity in Fig. 1d-e. Starting from lab frame tracked embryo centroids (section \ref{sec:SegmentTrack}), we first transformed the trajectories to the corotating frame (see section \ref{sec:corot} for more details). Individual embryo velocities were calculated by taking a time derivative of the tracked embryo positions, and then applying a 6-frame (1-minute) moving-window smoothing to reduce noise. From this corotating velocity field, we observed a region of synchronized velocity oscillation that grows, stabilizes, and then shrinks over time. During the stabilized oscillatory state, we manually cropped out a circular region of synchronized oscillation (with a diameter around 25 embryos for the exemplary experiment in Fig. 1). The resulting particle-based velocity dynamics is plotted in Fig. 1c, where the arrows represent particle-based velocity, and the particles are colored by velocity orientation (with respect to the horizontal x-axis). We then averaged the velocity field over the circular region, and plotted the velocity magnitude in the $x$-and $y$-directions in Fig. 1d.

\begin{figure}[h!]
  \centering
\includegraphics[width=\linewidth]{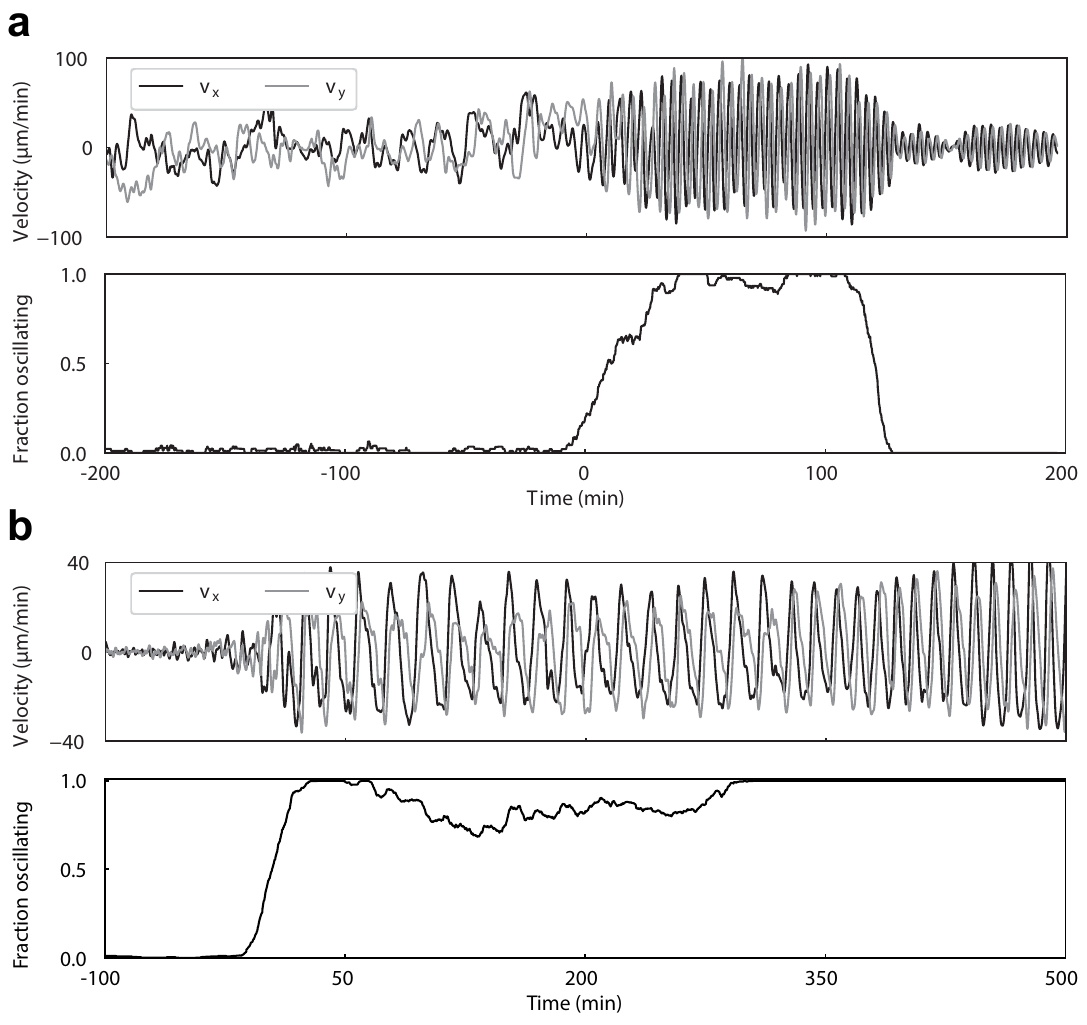}
    \caption{\textbf{ Velocity oscillation and fraction of oscillating embryos for two different spontaneous oscillation experiments.} The two experiments both exhibit stable velocity oscillations, persisting on the scale of hours. The 0 min. mark denotes the onset of oscillations.}
\label{figSI:VelOscFracB1B21}
\end{figure}

To determine whether an embryo occupied the oscillatory or fluctuating state as in Fig. 1e-f of the main text, we thresholded its oscillation amplitude. Based on our bistability hypothesis in tilt angle (eqn. \ref{eq:S4}), embryos can stabilize in either an upright or tilted precessing state; in our theoretical model (eqn. \ref{eq:S4}), this bistability is encoded in two stable fixed points: $\theta = 0$ (embryo upright / collective fluctuating state), and $\theta = \theta_0$ (embryo precession / collective oscillatory state), separated by an unstable fixed point $\theta = \theta_c$. Taking inspiration from this setup, identification of a characteristic velocity amplitude associated with $\theta_c$, $v_{\theta_c}$, would allow us to then identify whether an embryo is in the oscillatory or fluctuating state based on whether its measured velocity oscillation amplitude $v > v_{\theta_c}$ (oscillatory) or $v < v_{\theta_c}$ (fluctuating). We developed two methods to independently extract this unstable velocity oscillation amplitude, as described below.
 
\begin{figure}[h!]
  \centering
\includegraphics[width=\linewidth]{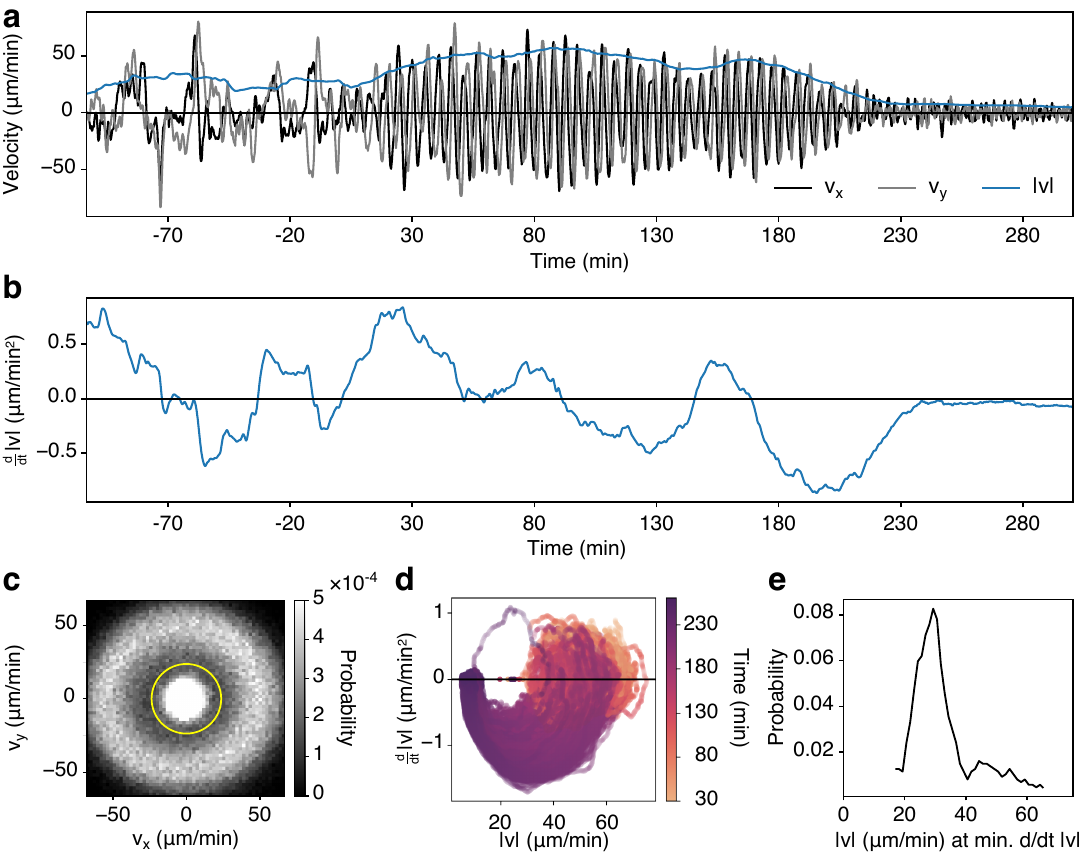}
    \caption{\textbf{Determination of velocity oscillation amplitude threshold for discriminating between oscillating and non-oscillating embryos.} \textbf{(a)} Time series of the velocity components (black, gray) for a representative single embryo trajectory. The blue line is the velocity oscillation amplitude averaged over five oscillation periods. \textbf{(b)} Rate of change of the velocity oscillation amplitude for a representative single embryo trajectory. \textbf{(c)} Probability density map of embryos with velocity components $v_x$ and $v_y$. The bright region near zero corresponds to embryos in the fluctuating state. The bright ring around the outer edge of the plot corresponds to embryos in the oscillatory state, and the yellow circle corresponds to the extracted velocity oscillation amplitude threshold. \textbf{(d)} Rate of change of the velocity oscillation amplitude plotted against the velocity oscillation amplitude for individual embryo trajectories. Embryos either relax to the fluctuating (low $|\mathbf{v}|$) or oscillatory (high $|\mathbf{v}|$) state, or transition from the oscillatory state to the fluctuating state. \textbf{(e)} Probability distribution of the velocity oscillation amplitude at the minimum rate of change of the velocity oscillation amplitude. The location of the peak of the distribution corresponds to the velocity oscillation amplitude threshold.}
  \label{figSI:VelBistable}
\end{figure}

The first method rests on the expectation that in a bistable system, embryos are more likely to occupy one of the two stable states than the unstable state corresponding to $v_{\theta_c}$ - that is, either fluctuating about $v=0$ in the fluctuating state or oscillating with some relatively constant  $v_{\theta_0}$ in the oscillatory state. We as such were motivated to examine the embryos' probability distribution in velocity phase space after stable LCC formation (fig. \ref{figSI:VelBistable}c), and identified the lowest probability velocity oscillation amplitude with $v_{\theta_c}$. For the exemplary experiment in Fig. 1 of the main text, $v_{\theta_c} = 23.7 \pm 10.7$ ($\mu$m/min).

The second method to determine the speed $v_{\theta_c}$ corresponding to the unstable tilt angle hypothesizes that the system will move faster away from the its unstable fixed point at $v=v_{\theta_c}$ than the two stable fixed points at $v=0$ and $v=v_{\theta_0}$. In other words, if the system transitions away from $v_{\theta_c}$ the fastest, we can identify this unstable velocity with the point of largest acceleration. To this end, we first determined the oscillation period for a given experiment from the average power spectrum of long-lived embryo trajectories ($0.0033 \text{min.}^{-1}$ for the exemplary experiment in Fig. 1). We then calculated the oscillation envelope by smoothing each embryo's instantaneous velocity over five oscillation periods (blue line in fig. \ref{figSI:VelBistable}a). We later calculated the acceleration as the discrete time derivative of the velocity oscillation amplitude, smoothing over six frames (one minute) to reduce noise at short time scales (fig. \ref{figSI:VelBistable}b). We then plotted the acceleration against the velocity oscillation envelope amplitude, shown in fig. \ref{figSI:VelBistable}d with individual embryo trajectories overlaid. The corresponding probability distribution of embryo oscillation amplitudes at minima in their accelerations (fig. \ref{figSI:VelBistable}e) allowed for precise quantification of the speed corresponding to maximum acceleration, $v_{\theta_c} = 29.5 \pm 5.0$ ($\mu$m/min) for the exemplary experiment in Fig. 1.

As these two methods yield consistent results for $v_{\theta_c}$, we used the value obtained from the simpler first method.

\subsection{Trajectory gap-filling}

The gaps in the tracked embryo trajectories are filled using interpolation on the trajectory using the MATLAB function \texttt{interp1} with the \texttt{pchip} option, which uses shape preserving cubic spline interpolation. Extrapolation at the end points is used if the number of missing frames at the end point is less than the memory parameter in the segmentation. 

\subsection{Smoothed displacement field from gap-filled trajectories}
\label{sec:smDisp}

To calculate a smooth displacement field, we apply a spatial Gaussian filter to regions where trajectories with long tracks exist. The region was identified by calculating the alpha-shape (using MATLAB \texttt{alphaShape} objects) of all segmented points along each trajectory tracked over the entire time window. The radius parameter is set to be approximately one embryo diameter, where the embryo diameters are independently determined from the first peak of the radial distribution function $g(r)$ for each dataset. Only particles that lie inside the alpha shape at a given frame are used in the coarse-graining. 

Non-smooth Fourier coefficients are first calculated using the MATLAB function
\texttt{nufftn} on the segmented point frame by frame. This results in a set of coefficients, 
\begin{subequations}
    \begin{align}
        \Tilde{u}^x_{k, l}(t) &= \sum_{n = 1}^{N(t)} u^x_n e^{2\pi i \left(\frac{k}{L_x} x_n(t) + \frac{l}{L_y} y_n(t) \right)} \\
        \Tilde{u}^y_{k, l}(t) &= \sum_{n = 1}^{N(t)} u^y_n e^{2\pi i \left(\frac{k}{L_x} x_n(t) + \frac{l}{L_y} y_n(t) \right)}
    \end{align}
\end{subequations}
where $(x_n(t), y_n(t))$ are the positions of the segmented points at time $t$, $N(t)$ is the number of points segmented at time $t$, $L_x$ and $L_y$ are the respective sizes of the imaging domain. The segmented displacement field $(u^x_n, u^y_n) = (x_n - \bar{x}, y_n - \bar{y})$ is calculated by subtracting the embryos trajectory median $(\bar{x}, \bar{y})$ from the segmented position. Using the median is robust to large changes in embryo position due to tracking or segmentation issues. To produce smooth displacement fields the resulting Fourier coefficients are multiplied by a Gaussian filter, $\exp(-\sigma (k^2/L_x^2 + l^2/L_y^2))$ with $\sigma = 3000 \,\text{px}^2$. Only coefficients that are above machine precision after the filtering are retained. Real space fields are then evaluated on a uniform grid using the fast Fourier transform \texttt{fft2} in MATLAB. 

Finally the smooth real space fields are cropped to pixels that lie inside the original alpha shape of the long particle trajectories. Note that this crops out the periodic boundary conditions assumed by the FFT.

\subsection{Lab frame to corotating frame}
\label{sec:corot}

Given a set of $N$ particle trajectories, $\{(x_n(t), y_n(t))\}_{n = 1}^N$ stacked into vectors $\mathbf{x}$ and $\mathbf{y}$, we seek to find the best fit rigid body rotation that minimizes the loss function
\begin{subequations} \label{eq:rotation_loss}
\begin{align}
    \mathcal{L}\left(\mathbf{x}_0, \mathbf{y}_0, \omega_R, c_x, c_y\right) &= \frac{1}{M}\sum_{m = 1}^M \left\lVert \mathbf{x}_0 \cos(\omega_R t_m) - \mathbf{y}_0 \sin(\omega_R t_m) + c_x \mathbf{1}- \mathbf{x}(t_m)\right\rVert_2^2 \nonumber \\
    &\hspace*{6em} + \frac{1}{M}\sum_{m = 1}^M \left\lVert \mathbf{x}_0 \sin(\omega_R t_m) + \mathbf{y}_0 \cos(\omega_R t_m) + c_y \mathbf{1}- \mathbf{y}(t_m)\right\rVert_2^2 \\
    M \mathcal{L} &=\sum_{m = 1}^M \mathbf{x}_0^\top \mathbf{x}_0 + \mathbf{y}_0^\top \mathbf{y}_0 + 2c_x \mathbf{x}_0^\top \mathbf{1} \cos(\omega_R t_m) - 2 \mathbf{x}_0^\top \mathbf{x} \cos(\omega_R t_m) \nonumber \\ 
    &\hspace{3em} - 2c_x \mathbf{y}_0^\top \mathbf{1} \sin(\omega_R t_m)  
    + 2\mathbf{y}_0^\top \mathbf{x} \sin(\omega_R t_m) + Nc_x^2 - 2c_x \mathbf{1}^\top \mathbf{x} + \mathbf{x}^\top\mathbf{x} \nonumber 
    \\ 
    &\hspace{3em} + 2c_y \mathbf{x}_0^\top \mathbf{1} \sin(\omega_R t_m) \nonumber  - 2\mathbf{x}_0^\top \mathbf{y} \sin(\omega_R t_m) + 2c_y \mathbf{y}_0^\top \mathbf{1} \cos(\omega_R t_m) \nonumber 
    \\ &\hspace{3em} - 2\mathbf{y}_0^\top \mathbf{y} \cos(\omega_R t_m) - 2c_y \mathbf{1}^\top \mathbf{y}  + \mathbf{y}^\top \mathbf{y} + Nc_y^2.
\end{align}
\end{subequations}
The rotation is parameterized by the frequency $\omega_R$, center of the rotation $(c_x, c_y)$ and the best fit initial positions $\mathbf{x}_0$, $\mathbf{y}_0$. We optimize the initial positions since $\mathbf{x}_0 \ne \mathbf{x}(0)$ if there are fluctuations on top of the rotation. In all the parameters except for $\omega_R$ the loss function is quadratic, we can, therefore, use a variable projection approach to convert the optimization into a one-dimensional problem. First we fix $\omega_R$ and solve the resulting optimization for $\mathbf{x}_0$, $\mathbf{y}_0$, $c_x$, and $c_y$. We will use $\langle \cdot \rangle$ to represent time averages with respect to the time samples and $\bar{\cdot}$ to represent averages over particles. 
\begin{subequations} \label{eq:rotation_loss_deriv}
\begin{align}
\frac{1}{2}\frac{\partial \mathcal{L}}{\partial \mathbf{x}_0} &= \mathbf{x}_0 + c_x \mathbf{1} \langle\cos(\omega_R t)\rangle - \langle\mathbf{x}\cos(\omega_R t)\rangle + c_y \mathbf{1} \langle\sin(\omega_R t)\rangle - \langle\mathbf{y} \sin(\omega_R t)\rangle \label{eq:rotation_loss_deriv_x0}\\ 
\frac{1}{2}\frac{\partial \mathcal{L}}{\partial \mathbf{y}_0} &= \mathbf{y}_0 - c_x \mathbf{1} \langle\sin(\omega_R t)\rangle + \langle\mathbf{x}\sin(\omega_R t)\rangle + c_y \mathbf{1} \langle\cos(\omega_R t)\rangle - \langle\mathbf{y} \cos(\omega_R t)\rangle \label{eq:rotation_loss_deriv_y0}\\ 
\frac{1}{2}\frac{\partial \mathcal{L}}{\partial c_x} &= N \bar{\mathbf{x}}_0 \langle \cos(\omega_R t)\rangle - N\bar{\mathbf{y}}_0 \langle \sin(\omega_R t)\rangle + N c_x - N\langle\bar{\mathbf{x}}\rangle \\
\frac{1}{2}\frac{\partial \mathcal{L}}{\partial c_y} &= N\bar{\mathbf{x}}_0 \langle \sin(\omega_R t)\rangle + N\bar{\mathbf{y}}_0 \langle \cos(\omega_R t)\rangle + N c_y - N\langle\bar{\mathbf{y}}\rangle
\end{align}
\end{subequations}
We, therefore, get a simple linear system by setting eqn. \ref{eq:rotation_loss_deriv} to zero to solve for the parameters:
\begin{multline} \label{eq:rotation_linsys}
    \begin{pmatrix} 
    I_{N} & 0_N & \mathbf{1} \langle\cos(\omega_R t)\rangle & \mathbf{1} \langle\sin(\omega_R t)\rangle \\ 
    0_{N} & I_N & -\mathbf{1} \langle\sin(\omega_R t)\rangle & \mathbf{1} \langle\cos(\omega_R t)\rangle \\ 
    \mathbf{1}^\top \langle\cos(\omega_R t)\rangle & - \mathbf{1}^\top \langle\sin(\omega_R t)\rangle & N & 0 \\ 
    \mathbf{1}^\top \langle\sin(\omega_R t)\rangle & \mathbf{1}^\top \langle\cos(\omega_R t)\rangle & 0 & N
    \end{pmatrix}
    \begin{pmatrix} 
    \mathbf{x}_0 \\ \mathbf{y}_0 \\ c_x \\ c_y
    \end{pmatrix}
    \\ = 
    \begin{pmatrix} 
    \langle \mathbf{x} \cos(\omega_R t) \rangle + \langle \mathbf{y} \sin(\omega_R t) \rangle \\ 
    -\langle \mathbf{x} \sin(\omega_R t) \rangle + \langle \mathbf{y} \cos(\omega_R t) \rangle \\
    N\langle\bar{\mathbf{x}}\rangle \\ 
    N\langle\bar{\mathbf{y}}\rangle
    \end{pmatrix}
\end{multline}
Substituting the solution of equation\eqref{eq:rotation_linsys} into \eqref{eq:rotation_loss} yields a single function $\mathcal{L}(\omega_R)$ that can be minimized using standard single variable minimization techniques. Given values of $\omega_R$, $c_x$ and $c_y$ and a trajectory $(x_i(t), y_i(t))$, we can find a suitable $(x_0, y_0)$ by solving the scalar optimization that results from equations \eqref{eq:rotation_loss_deriv_x0}~and \ref{eq:rotation_loss_deriv_y0}, with all other parameters fixed.

We need to consider the special case when $\omega_R = 0$,
\begin{align*}
    \frac{1}{2}\frac{\partial \mathcal{L}}{\partial \mathbf{x}_0} &= \mathbf{x}_0 + c_x \mathbf{1}  - \langle\mathbf{x}\rangle \\ 
    \frac{1}{2}\frac{\partial \mathcal{L}}{\partial \mathbf{y}_0} &= \mathbf{y}_0 + c_y \mathbf{1} - \langle\mathbf{y} \rangle \\ 
    \frac{1}{2}\frac{\partial \mathcal{L}}{\partial c_x} &= N \bar{\mathbf{x}}_0 + N c_x - N\langle\bar{\mathbf{x}}\rangle \\
    \frac{1}{2}\frac{\partial \mathcal{L}}{\partial c_y} &= N\bar{\mathbf{y}}_0 + N c_y - N\langle\bar{\mathbf{y}}\rangle
\end{align*}
which has the solution, $\mathbf{x_0} = \langle\mathbf{x}\rangle - c_x \mathbf{1}$ and $\mathbf{x_0} = \langle\mathbf{x}\rangle - c_x \mathbf{1}$. Substituting into the derivatives with respect to $(c_x, c_y)$
\begin{align*}
    \mathbf{1}^\top \mathbf{x_0} = N \bar{\mathbf{x}}_0 = \mathbf{1}^\top (\langle\mathbf{x}\rangle - c_x \mathbf{1}) = N\langle\bar{\mathbf{x}}\rangle - N c_x \\ 
    \mathbf{1}^\top \mathbf{y_0} = N \bar{\mathbf{y}}_0 = \mathbf{1}^\top (\langle\mathbf{y}\rangle - c_y \mathbf{1}) = N\langle\bar{\mathbf{y}}\rangle - N c_y
\end{align*}
we immediately see that the derivatives are set to $0$ for any choice of $c_x$ and $c_y$. We therefore choose to set $c_x = \langle\bar{\mathbf{x}}\rangle$ and  $c_y = \langle\bar{\mathbf{y}}\rangle$.

The corotating displacement field is, therefore, given by \begin{subequations}
\begin{align}
    \mathbf{u}_x(t) &= (\mathbf{x}(t) - c_x)\cos(\omega_R t) + (\mathbf{y}(t) - c_y)\sin(\omega_R t) - \mathbf{x}_0 \nonumber \\
    \mathbf{u}_y(t) &= -(\mathbf{x}(t) - c_x)\sin(\omega_R t) + (\mathbf{y}(t) - c_y)\cos(\omega_R t) - \mathbf{y}_0 \nonumber\\
    \mathbf{u} &= (\mathbf{z} - c_z)e^{-i\omega_R t} - \mathbf{z}_0 \label{eq:urot}\\ 
    \mathbf{z} &= (\mathbf{u} + \mathbf{z}_0) e^{i\omega_R t} + c_z \label{eq:zrot}
\end{align}
\end{subequations}

\subsection{Dispersion relation analysis}
\label{SI:dispersion}

\subsubsection{Dynamic mode decomposition}
\label{SI:DMD}
Dynamic mode decomposition assumes a linear time-invariant dynamics of the system and models the displacement field as $\dot{\mathbf{u}} = M \mathbf{u}$. An alternative way of writing this assumption is as an expansion of the dynamics into a set of modes that have exponential time variation,  
\begin{equation} \label{eq:udmd}
    \mathbf{u} = \sum_{n = 0}^N \boldsymbol{\phi}_n e^{i\omega_n t} = \sum_{n = 0}^N \boldsymbol{\phi}_n e^{i\nu_n t} e^{-\lambda_n t}
\end{equation}
where $i\omega_n = -\lambda_n + i\nu_n$ are the complex eigenvalues of $M$, and $\boldsymbol{\phi}_n$ are the corresponding eigenvectors. 
The corresponding expansion for embryo positions, found by substituting equation \ref{eq:udmd} into equation \ref{eq:zrot}, is given by
\begin{equation}
    \mathbf{z} = c_z\mathbf{1} + \mathbf{z}_0 e^{i\omega_R t} + \sum_{n = 0}^N \boldsymbol{\phi}_n e^{i(\nu_n + \omega_R) t} e^{-\lambda_n t}.
\end{equation}
Following~\cite{askham2018variable}, to fit expansions of the form given in eqn. \ref{eq:udmd} we minimize the loss function
\begin{subequations} \label{eq:DMD_loss_full}
\begin{align} 
    \mathcal{L}_{DMD}\left(\{(\nu_n, \lambda_n, \boldsymbol{\phi}_n)\}_{n = 1}^N; \{\mathbf{u}(t_m)\}_{m = 1}^M\right) &= \sum_{m = 1}^M \left\lVert \mathbf{u}(t_m) - \sum_{n = 0}^N  \boldsymbol{\phi}_n e^{i\nu_n t} e^{-\lambda_n t} \right\rVert^2 \\
    &= \sum_{m = 1}^M \lVert \mathbf{u}(t_m) \rVert^2 + \sum_{n = 0}^N \sum_{k = 0}^N \boldsymbol{\phi}_n^\dagger \boldsymbol{\phi}_k e^{i (\nu_k - \nu_n) t} e^{-(\lambda_n + \lambda_k)t} \nonumber\\
    & \qquad - \sum_{n = 0}^N \mathbf{u}(t_m)^\dagger \boldsymbol{\phi}_n e^{i\nu_n t} e^{-\lambda_n t} + \boldsymbol{\phi}_n^\dagger \mathbf{u}(t_m) e^{-i\nu_n t} e^{-\lambda_n t}
\end{align}
\end{subequations}
over $\{(\nu_n, \lambda_n, \boldsymbol{\phi}_n)\}_{n = 1}^N$. For the dynamics to be stable we require $\lambda_n \ge 0$ which we add as a constraint to the optimization.
Again we notice that for fixed $\nu_n$ and $\lambda_n$ the problem is quadratic in $\boldsymbol{\phi}_n$. 

The optimal $\boldsymbol{\phi}_n$ is given by the solution of the least squares problem,
\begin{subequations}
    \begin{equation} \label{eq:phi_leastsquares}
    \Phi T = U \text{ where } \Phi_{nk} = (\boldsymbol{\phi}_k)_n, T_{k, m} = e^{i \nu_k t_m} e^{-\lambda_k t_m} \text{ and } U_{n, m} = (\mathbf{u}(t_m))_n
\end{equation}
which has the solution, 
\begin{equation}
    \Phi = UT^\dagger (T T^\dagger)^{-1}. \label{eq:phi_solution}
\end{equation}
In practice $(T T^\dagger)^{-1}$ can be ill-conditioned if the $\omega_n$'s become close to each other, which can happen during optimization. To overcome this we add a small amount of regularization $\gamma \lVert \Phi \rVert^2$ to the least squares problem, yielding the regularized solution
\begin{equation}
    \Phi = UT^\dagger (T T^\dagger + \gamma I)^{-1}. \label{eq:phi_solution_reg}
\end{equation}
where $\gamma$ is chosen to be small. 
\end{subequations}
Using the solution (eqn. \ref{eq:phi_solution_reg}) in eqn. \ref{eq:DMD_loss_full} yields a non-linear optimization over the $\nu_n$ and $\lambda_n$, 
\begin{equation} \label{eq:DMD_loss_varproj}
\min_{\nu_n, \lambda_n} \mathcal{L}_{DMD} 
\end{equation}
As has been noted before, the optimization in eqn. \ref{eq:DMD_loss_varproj} is highly non-convex and gradient based optimizations are sensitive to initial conditions. We, therefore, start by using a gradient-free optimization, particle swarm~\cite{kennedy1995particle}, to get a good initialization for a later gradient based optimization~\cite{askham2018variable}.

\begin{figure}[h!] \centering\includegraphics[width=0.6\linewidth]{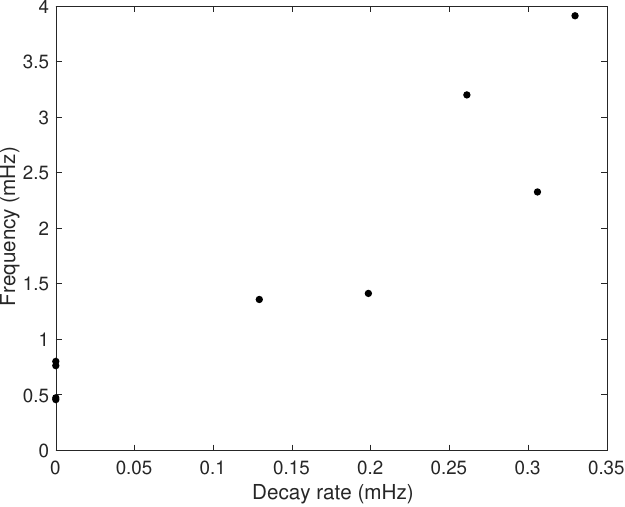}
    \caption{\textbf{Relation between oscillation frequency and decay rate in fluctuating state dynamical modes.} In the fluctuating state, only acoustic modes are present. Similarly to Fig. 2b, the decay rates of these modes are an order of magnitude smaller than their corresponding imaginary frequencies, agreeing with previous estimates of the ratio between odd and even elastic moduli \cite{tan2022odd}.}
  \label{fig:staticfreqs}
\end{figure}

As discussed in the main text, we see that in the fluctuating state, the real part of the DMD frequencies is roughly an order of magnitude larger than the imaginary part (see fig. \ref{fig:staticfreqs}).

\subsubsection{Determining optimal number of modes for reconstruction}
\label{SI:Nmode_DMD_reco}
\begin{figure}[h!] \centering\includegraphics[width=0.6\linewidth]{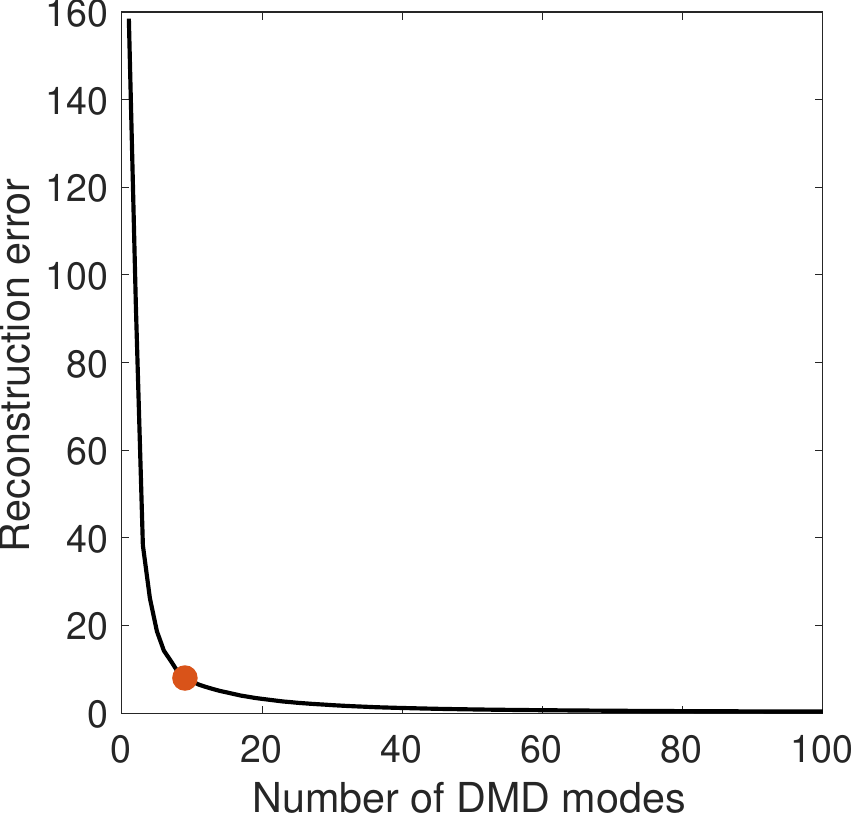}
    \caption{\textbf{Optimal number of modes for DMD analysis.} Optimal loss (eqn. \ref{eq:DMD_loss_varproj}) as a function of number of modes. We choose a number of modes in the elbow of the plot at $N = 9$, as indicated by the orange dot.}
  \label{fig:DMDLoss}
\end{figure}

If the dynamics are truly linear, then an exact reconstruction is possible when the number of modes is equal to the number of embryos. In practice, however, we find that the dynamics are approximately low rank and we can choose $N$ to be smaller. To determine the number of modes, we plot the optimization loss $\mathcal{L}_{\mathrm{DMD}}$ as a function of the number of modes~(see~fig.~\ref{fig:DMDLoss}). We choose a value for $N$ in the elbow of the curve which represents a trade off between reconstruction error and number of modes. In the remaining analysis, we choose $N = 9$. 

\subsubsection{Extracting dispersion relations from Dynamic Mode Decomposition}

Given an expansion of the form in eqn. \ref{eq:udmd}, we can extract a dispersion relation by relating the spatial component of each dynamical mode ($\boldsymbol{\phi}_n$) with a wavenumber. Given a field $\phi(\mathbf{x})$ over a domain $D$ we can assign a spatial wave vector using the formula
\begin{equation} \label{eq:freq_def_cont}
    |\mathbf{q}|^2 = - \left(\int_D \mathrm{d}\mathbf{x} \, \phi(\mathbf{x})^* \phi(\mathbf{x})\right)^{-1} \int_D \mathrm{d}\mathbf{x} \, \phi(\mathbf{x})^* \nabla^2 \phi(\mathbf{x}).
\end{equation}
In the case of a plane wave $\phi(\mathbf{x}) = e^{i \boldsymbol{\xi} \cdot \mathbf{x}}$, then $|\mathbf{q}|^2 = |\boldsymbol{\xi}|^2$, exactly the expected wave vector for pure harmonic spatial variation. 

Recall that the DMD modes are defined on the grid defined by the best-fit initial positions of the global rotation, $(\mathbf{x}_0, \mathbf{y}_0)$. We therefore discretize eqn. \ref{eq:freq_def_cont} on the $(\mathbf{x}_0, \mathbf{y}_0)$. First we define a graph over the points using an alpha shape triangulation with radius equal to embryo diameter or approximate grid spacing and all holes inside the shape filled. Positions outside of the largest connected component are removed as outliers. From the adjacency matrix of the triangulation, we calculate the FEM discrete Laplacian~\cite{reuter2009discrete}, which approximates the continuous Laplacian under grid refinement. This produces a matrix $L$ and we approximate the integral (eqn. \ref{eq:freq_def_cont}) with the inner product 
\begin{equation} \label{eq:freq_def_discrete}
    |\mathbf{q}_n|^2 = - \frac{\boldsymbol{\phi_n}^\dagger L \boldsymbol{\phi_n}}{\boldsymbol{\phi_n}^\dagger .\boldsymbol{\phi_n}}.
\end{equation}

\subsection{Mode-mode correlations}
\subsubsection{Pearson correlation}
The dynamic modes $\mathbf{u}_{n}(t) = \boldsymbol{\phi_{n}}e^{i\nu_{n}t}e^{-\lambda_{n}t}$ obtained from DMD can be decomposed into a real component $\mathbf u_{n,\mathcal{R}}(t)$ and an imaginary component $\mathbf u_{n,\mathcal{I}}(t)$.  The mean $\bar{\mathbf u}_{\mathcal{R},n} = \frac{1}{T}\sum_{t=1}^{T}\mathbf u_{\mathcal{R},n}$ and variance $\sigma^{2}_{\mathcal{R},n} = \frac{1}{T-1}\sum_{t=1}^{T}(\mathbf u_{\mathcal{R},n}-\bar{\mathbf u}_{\mathcal{R},n})^{2}$ can be calculated and used to define the normalized correlation matrix $\textbf{C}^{\mathcal{R},\mathcal{R}}_{n,m}$:

\begin{equation}\textbf{C}^{\mathcal{R},\mathcal{R}}_{n,m} = \frac{1}{T} \sum_{t = 1}^{T}\frac{(\mathbf u_{n,\mathcal{R}}(t)-\bar{\mathbf u}_{\mathcal{R},n})(\mathbf u_{m,\mathcal{R}}(t)-\bar{\mathbf u}_{\mathcal{R},m})}{\sigma_{\mathcal{R},n}\sigma_{\mathcal{R},m}}  \end{equation}

Analogous quantities can be used to define the imaginary-imaginary correlation matrix $\textbf{C}^{\mathcal{I},\mathcal{I}}_{n,m}$ and the imaginary-real cross correlation matrix $\textbf{C}^{\mathcal{I},\mathcal{R}}_{n,m}$.

\subsubsection{Spatial mode correlations}

The dynamic mode decomposition enables a functional separation of spatial and temporal scales in the evolution of the system dynamics. To quantify the spatial similarity between modes, we calculate the matrix $\boldsymbol{\Phi}_{nm} = \left<\boldsymbol{\phi}_{m}\cdot\boldsymbol{\phi}_{n}\right>$, where $\cdot$ is the dot product between eigenvectors. We normalize this matrix as described in the previous section so that the diagonal terms are of unit magnitude.

\subsubsection{Normalized correlation matrix and angular momentum tensor}
\label{ssec:angmomtensor}

As in  \cite{bacanu2023inferring} The modes can also be used to define the antisymmetrized correlation functions
\begin{equation}\mathcal{L}^{\mathcal{R},\mathcal{R}}_{n,m}(\tau) = \left<\mathbf u_{n}(t+\tau)\mathbf u_{m}(t)-\mathbf u_{m}(t+\tau)\mathbf u_{n}(t)\right>_{t}\end{equation}

The antisymmetrized correlation functions $\mathcal{L}_{n,m}^{\mathcal{R},\mathcal{R}}(\tau)/\sigma_{\mathcal{L}^{\mathcal{R}\mathcal{R}}_{nm}}(\tau)$ can be normalized by the corresponding standard deviation at each time point, with the variance being calculated as
$$\sigma^{2}_{\mathcal{L}^{\mathcal{R}\mathcal{R}}_{nm}}(\tau) = \left<(\mathbf u_{n}(t+\tau)\mathbf u_{m}(t)-\mathbf u_{m}(t+\tau)\mathbf u_{n}(t)-\mathcal{L}^{\mathcal{R},\mathcal{R}}_{n,m}(\tau))^{2}\right>_{t}. $$

As in \cite{bacanu2023inferring}, the sign and magnitude of the angular momentum is directly related to nonequilibrium activity and steady-state cycles in the subspace spanned by variables used to construct the angular momentum function.  Specifically, the cycling direction and the magnitude is directly related to the sign and area of the angular momentum.
\begin{figure}[h!]
  \centering
  \includegraphics[width=\linewidth]{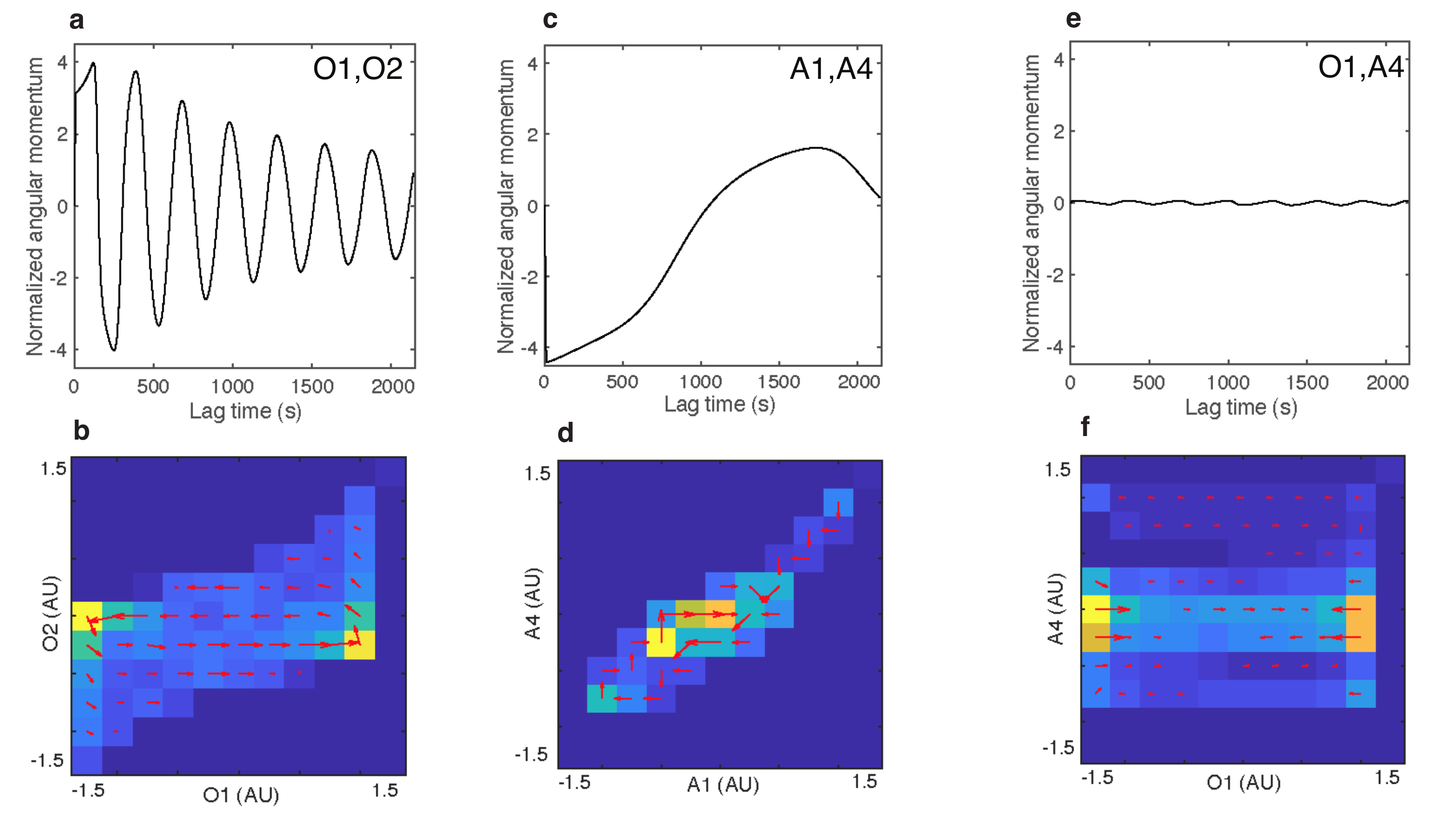}
    \caption{\textbf{Angular momentum traces.} Time series of the asymmetric correlation function between two optical modes \textbf{(a)}, two acoustic modes \textbf{(c)}, and one optical and one acoustic mode \textbf{(e)}. The angular momentum of the two optical modes exhibits a larger magnitude than that of the two acoustic modes, which in turn is larger than that of the optical and acoustic mode pair.  Furthermore, the two optical modes exhibit the fastest oscillation dynamics. The mode pair correlations' corresponding phase space cycles \textbf{(b, d, f)} encode further information about the nature of the nonequilibrium dynamics within the system.  The sign of the angular momentum corresponds to the cycling direction.  Very little cycling is seen in \textbf{(f)} compared to \textbf{(b)}, hence the angular momentum also quantifies cycling magnitude.}
  \label{fig:angmomcycles}
\end{figure}

\subsubsection{Dendrogram}
To quantify the similarity between each mode pair in the correlation matrix $\textbf{C}^{\mathcal{R},\mathcal{R}}_{n,m}$, a standard agglomerative clustering algorithm is used (\texttt{Hierarchical Clustering} in MATLAB). Briefly, the Euclidean distance between each pair of rows and columns is calculated. This encodes the linkage mapping, a metric that is used to permute the indices of the correlation matrix to specify if certain mode behaviors fall into discrete clusters.  

\subsection{Analysis of velocity and strain cycles}\label{sec:StrainCycle}

This section outlines how we obtain the velocity and strain cycles shown in Fig. 3c-d of the main text. We start by interpolating DMD reconstructions of the displacement field in the corotating frame of the LCC's global oscillation (see sections \ref{sec:corot}-\ref{SI:DMD}) to obtain a continuous field from which to calculate strain and velocity components. We interpolate the particle-based displacement field $\mathbf{u}(\mathbf{r},t)$ in the corotating frame to a rectangular grid with a spatial grid size of one embryo. The velocity field is calculated by taking a time derivative of the displacement field, $\mathbf{v}(\textbf{r},t) = \frac{d}{dt} \mathbf{u}(\textbf{r},t)$. To obtain the strain rate fields, we first calculate the components of strain field from displacement fields, following the convention used in \cite{scheibner2020odd}: (1) divergence: $u^0(\textbf{r},t) = u_{xx}+u_{yy} $; (2) curl: $u^1(\textbf{r},t) = u_{yx}-u_{xy} $; (3) shear 1: $u^2(\textbf{r},t) = u_{xx}-u_{yy} $; (4) shear 2: $u^3(\textbf{r},t) = u_{yx}+u_{xy} $, where the displacement gradient is denoted as $u_{ij} = \partial_i u_j$ and $i,j \in \{x,y\}$. 

Having defined our relevant fields, we can proceed to obtain probability density currents in velocity and strain phase space. We consider a time series of a pair of physical variables $\textbf{x} (t)$, where $\textbf{x}(t) = (v_x(t), v_y(t))^T$ for velocity phase space, and $\textbf{x}(t) = (u^1 (t), u^0 (t))^T$ for curl-divergence strain phase space. The time series (with $N_t$ total time points) is sampled at $\Delta t$ intervals. The variables are taken from all spatial regions from the reconstructed fields during the oscillatory state (about 80 minutes of stable oscillations for the example in Fig. 1 of the main text). We then apply kernel density estimation to estimate the continuum probability density $\hat{\rho} (\hat{\textbf{x}})$ and its corresponding probability currents $\hat{\textbf{j}} (\hat{\textbf{x}})$ from the observed time series \cite{tan2022odd,li2019quantifying,just2003nonequilibrium,bowman1997applied} (see section \ref{SI:irreversibility} for definitions of the probability density and current). The cycles themselves are visualizations of $\hat{\rho} (\hat{\textbf{x}})$ and $\hat{\textbf{j}} (\hat{\textbf{x}})$, which provide the cycle's phase space structure (color bar in Fig. 3c-d) and direction (overlaid arrows) respectively.

\begin{figure}[h!]
  \centering
  \includegraphics[width=\linewidth]{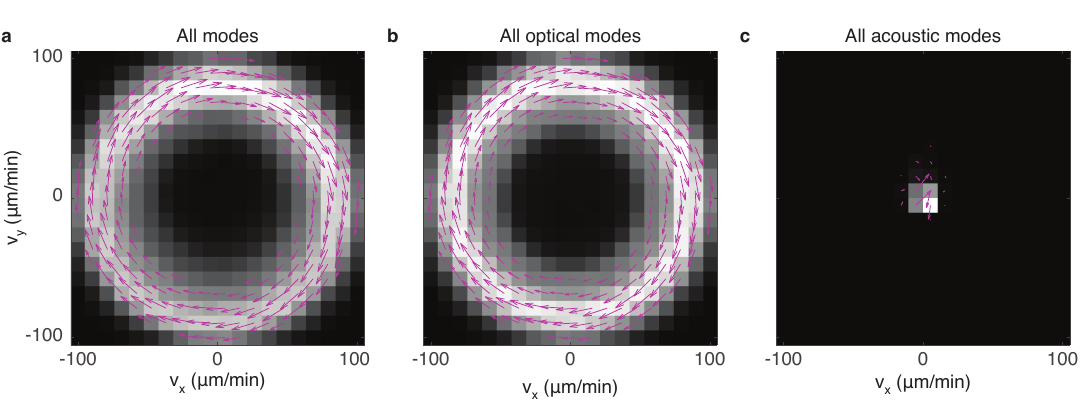}
    \caption{\textbf{Velocity phase space cycles reconstructed from DMD modes.} Phase space cycles in velocity space constructed from \textbf{(a)} all modes, \textbf{(b)} all optical modes, and \textbf{(c)} all acoustic modes. The heat map color-codes the estimated probability density $\hat{\rho}(\textbf(x))$, where white denotes a higher probability density. The arrows display the estimated probability current $\hat{\textbf{j}}(\textbf{x})$.}
  \label{fig:dmdVelCycle}
\end{figure}

Just as the DMD modes can be separated into optical and acoustic modes, so the displacement fields can be separately reconstructed using only one set of modes or the other to differentiate their relative contributions to the phase space cycles. The velocity field reconstruction from solely optical modes exhibits similar phase space cycles compared to the reconstruction from all DMD modes (fig. \ref{fig:dmdVelCycle}a-b). On the other hand, no clear velocity phase space cycle is observed in the velocity field reconstructed from solely acoustic modes (fig. \ref{fig:dmdVelCycle}c). These results are consistent with our expectations from our continuum theory (section \ref{sec2:1}), where velocity oscillations result from precession, an effect captured predominantly by the DMD optical modes.

\subsection{Tilt angle analysis}
\label{SI:tiltcalc}
\begin{figure}[h!]
  \centering
  \includegraphics[width=0.3\linewidth]{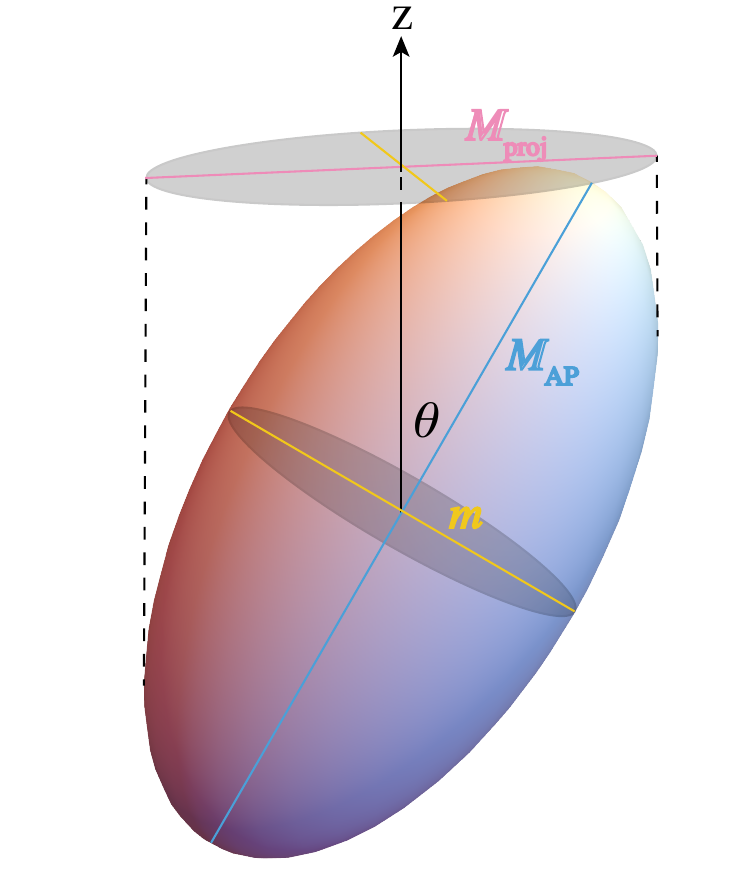}
  \includegraphics[width=0.5\linewidth]{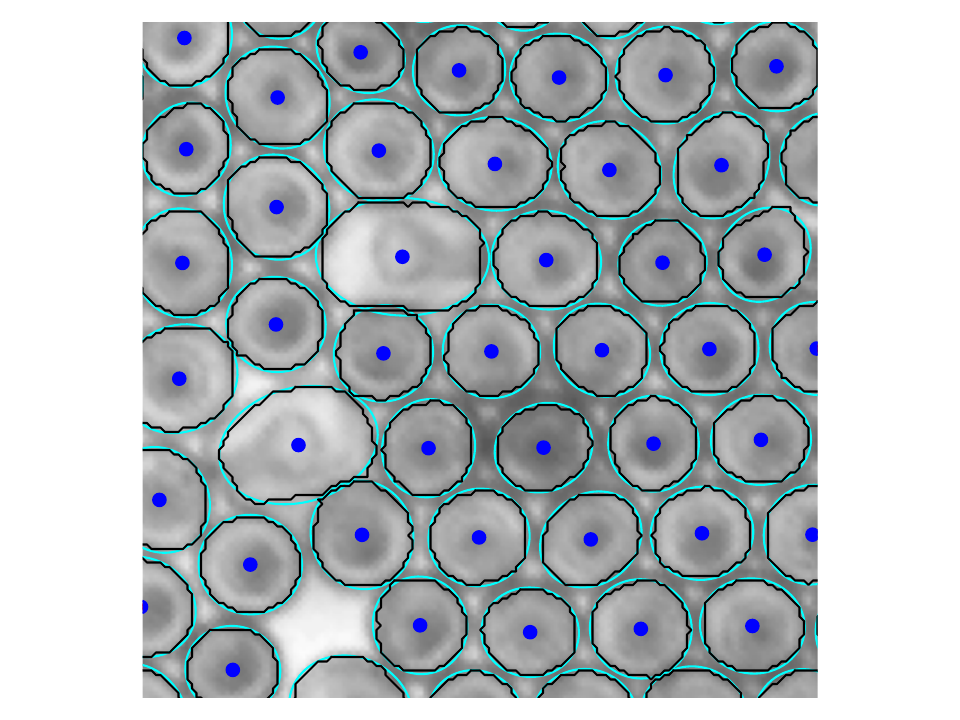}
    \caption{\textbf{Estimating tilt angle from ellipsoid projection and embryo contours.} (Left) Schematic of ellipsoid with tilt angle $\theta$ from vertical. (Right) microscope image of embryos (top-down) with raw image segmentation contours (black) and ellipse fits (green) overlaid; embryo centroids are marked with blue dots.}
  \label{fig:ellipsoid_axes}
\end{figure}

To estimate embryo tilt angles from 2D microscope images, we assume the embryos to be ellipsoids as in fig. \ref{fig:ellipsoid_axes}, i.e. to have ellipses as 2D projections. Such ellipses, examples of which are shown in fig. \ref{fig:ellipsoid_axes} (cyan), describe embryo contours similarly to those obtained directly from image segmentation (black) (see section \ref{sec:SegmentTrack} for details of image segmentation). 

Given the major ($M_{proj}$) and minor ($m$) axis lengths of the projected ellipses, and the anterior-posterior (AP) axis length ($M_{AP}$) of the embryo, we can estimate the tilt of each embryo from the vertical as
\begin{equation}
\label{eqn:tiltangle}
    \theta = \arccos \left(\sqrt{\frac{M_{AP}^2-M_{proj}^2}{M_{AP}^2-m^2}} \right)
\end{equation}
This equation is derived in the following subsection. As the AP axis length is not directly accessible from our 2D microscopy images, we estimate it as $M_{AP} = \alpha_{94} m$, where $\alpha_{94} \approx 1.4$ is the 94th percentile aspect ratio across all embryos in each experiment. We obtain ellipse major and minor axis lengths from image segmentation using \texttt{cellpose} and \texttt{opencv}. We also smooth $\theta$ in time over a period of 30 seconds to reduce noise. In Fig. 4, embryos outlined in black correspond to those with lifetimes shorter than this smoothing window (resulting in NaN values).

\begin{figure}[h!]
  \centering
  \includegraphics[width=0.9\linewidth]{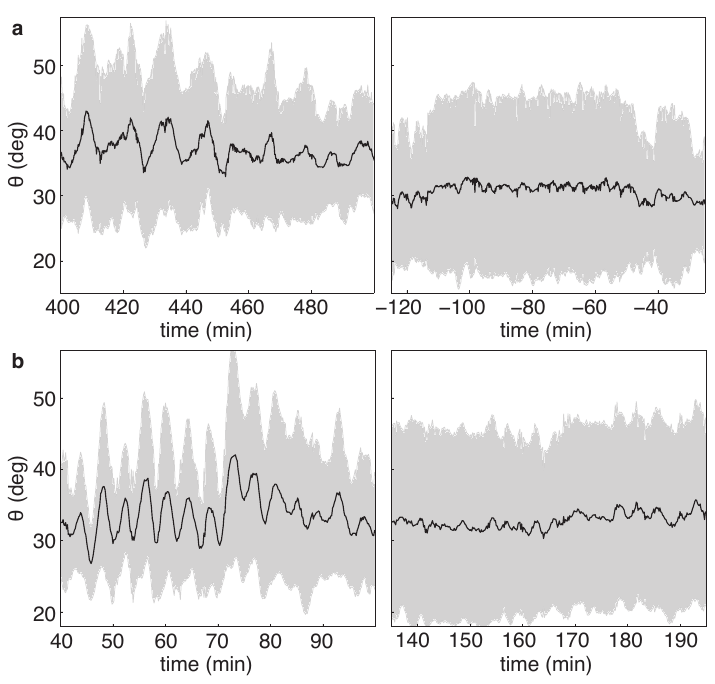}
    \caption{\textbf{Tilt angle oscillations in experimental replicates.} Tilt angle timelapses for two different replicates (\textbf{(a)} and \textbf{(b)}), averaging across a 120 px radius circle, comparing between time periods in which the system displays oscillations (\textbf{(a)} and \textbf{(b)}, left) and fluctuates in place (\textbf{(a)} and \textbf{(b)}, right). The 0 min mark denotes the onset of oscillations.}
  \label{fig:replicates_theta}
\end{figure}

Tilt analysis timelapses are shown for two additional replicates in fig. \ref{fig:replicates_theta}. As in Fig. 4 in the main text, they show robust signatures of oscillations easily distinguishable from their fluctuating state behaviors. The bottom replicate depicts a qualitatively higher average tilt angle in the oscillatory state, relative to the fluctuating state; the top one does not.

\subsubsection{Derivation of tilt angle equation}
In this section we derive eqn. \ref{eqn:tiltangle} for the tilt angle of a prolate spheroid with radius $r$ and height $2h$ from the vertical. Without loss of generality, take the vertical axis to be the $z$-axis and assume the ellipsoid is rotated about the y-axis so that it tilts away from the vertical by some angle $\theta$. Then the following ellipsoid equation holds:
\begin{equation}
    \frac{x^2+y^2}{r^2} + \frac{z^2}{h^2} = 1 = \frac{(\cos \theta x' - \sin \theta z')^2 + y'^{2}}{r^2} + \frac{(\sin \theta x' + \cos \theta z')^2}{h^2} \equiv F(\mathbf x'),
\end{equation}
where primes (') denote the rotated coordinates $(x',y',z') = \mathbf{r}'=R_{3\text{D}}(\theta)\mathbf{r}$. Our goal is to extract the projected major axis length $M_{proj} = 2x'(\theta; r,h)$ of the ellipsoid on the $x$-axis; this value corresponds to the major axis of the ellipse we extract from image segmentation of top-down embryos, described in the previous section. From there, we invert the expression for $x'$ ($M_{proj}$) to obtain the tilt angle $\theta(x';r,h)$ as a function of the ellipse parameters. 

To do so, we first project the ellipsoid onto a plane containing the $x$-axis, choosing the $xz$-plane ($\partial_y F = 0$) for simplicity. This ellipse obeys the equation
\begin{equation}
\label{eqn:ellipse_xz}
    \frac{(\cos \theta x' - \sin \theta z')^2}{r^2} + \frac{(\sin \theta x' + \cos \theta z')^2}{h^2} =1
\end{equation}
We now extremize the above equation to obtain the limits of the ellipse. Optimizing this equation with respect to $z'$ yields a maximal vertical ($z$) ellipse extent 
\begin{equation}
    z' = \Delta x',\quad \Delta = \frac{(h^2-r^2) \sin \theta \cos \theta}{h^2 \sin^2 \theta + r^2 \cos^2 \theta}
\end{equation}
We now substitute this expression back into eqn. \ref{eqn:ellipse_xz} to solve for $x'$, finding that
\begin{equation}
    x'^2 = h^2 \sin^2 \theta + r^2 \cos^2 \theta
\end{equation}
Since our derivation by construction has no component of the projected major axis along the $y$-direction (for a purely upright ellispoid rotated about the $y$-axis), the above equation gives us the major axis bounds for the $xy$-plane projection of the ellipsoid. Inverting it to solve for $\theta$,
\begin{equation}
    \theta = \arccos\left( \frac 12 \sqrt{\frac{h^2-x^2}{h^2-r^2}}\right)
\end{equation}
Identifying the embryo's AP axis length $M_{AP} = 2h$ and the minor axis length $m = 2r$, we plug in these relations to obtain
\begin{equation}
    \theta = \arccos\left( \sqrt{\frac{M_{AP}^2-M_{proj}^2}{M_{AP}^2-m^2}}\right)
\end{equation}
as desired.

\subsection{Divergence, rotation frequency, and phase analyses}
The following three subsections detail how we analyzed the divergence and rotation frequency fields shown in Fig. 4d-g, as well as the corresponding divergence-rotation frequency phase analyses in Fig. 4g (inset). 
\subsubsection{Divergence strain field analysis}
To resolve the divergence strain fields with particle-level resolution, we follow the procedure for particle-based extraction of strain fields in~\cite{tan2022odd}, which we briefly describe below. This method provides us with a direct avenue of comparison with other particle-based results, namely the rotation frequency analysis of the next section and the tilt angle analysis of the previous section, allowing us in particular to directly compare phase trajectories of the same embryos across different observables (divergence and rotation frequency) in section \ref{SI:phase}.

We consider a Delaunay triangulation of embryo centroids to identify nearest neighbors and define the ``local displacement field'' around a 6-coordinated embryo $i$ as 
\begin{equation}\label{eq:locDisp}
    \textbf{u}_i(j_n)=\textbf{r}_{j_n(i)}-\textbf{r}^\text{Tem}_{j_n(i)},
\end{equation}
which corresponds to the set of six displacement vectors $\textbf{u}_i(j_n)$ associated with the six nearest neighbors $j_1(i),...,j_6(i)$ of embryo $i$, located at positions $\textbf{r}_{j_n(i)}$. The positions $\textbf{r}^\text{Tem}_{j_n(i)}$ correspond to the corners of a template hexagon that is defined as follows. First, we calculate the lattice constant $\bar{a}^\text{Tem}$ as the mode of edge length distribution in Delaunay triangulation over all space and time. Next, we find the hexagonal bond orientation $\phi_i^\text{Tem}=\arg\langle\psi_6\rangle/6$ (mod~$\pi/3$), where local bond orientation order parameter $\psi_6$ is defined in \cite{tan2022odd,nelson1979}. Template positions surrounding each embryo $i$ are then given by the corners of a hexagon located~at
\begin{equation}\label{eq:Def_Tem}
    \mathbf{r}^\text{Tem}_{j_n(i)}=\mathbf{r}_{i}+\bar{a}^\text{Tem}\begin{bmatrix}
\cos\left(\frac{(n-1)\pi}{3}+\phi^\text{Tem}_i\right)\vspace{0.05cm}\\
\sin\left(\frac{(n-1)\pi}{3}+\phi^\text{Tem}_i\right)
\end{bmatrix}.
\end{equation}
Neighbor positions $\mathbf{r}_{j_n(i)}$ in equation~\ref{eq:locDisp} are indexed in counter-clockwise order and such that the bond angle between embryos $\mathbf{r}_{i}$ and $\mathbf{r}_{j_1(i)}$ is the one closest to $\phi^\text{Tem}_i$. Finally, we perform a linear regression (using MATLAB's~\verb+polyfitn+ function) and fit the 6 local displacement vectors~by
\begin{equation}\label{eq:Du}
\textbf{u}_{i,\text{fit}}(j_n)=\textbf{u}_0+\mathbf{S}(\mathbf{r}_i)^\top\cdot\textbf{r}_{j_n(i)}^\text{Tem},
\end{equation}
to determine a vector $\textbf{u}_0$ that captures any residual translations, while the fitted matrix
\begin{equation}\label{eq:Duf}
\mathbf{S}(\mathbf{r}_i)=\begin{pmatrix}
S_{11} & S_{12}\\
S_{21} & S_{22}
\end{pmatrix}
\end{equation}
approximates the local displacement gradient tensor $S_{ij}\approx\partial_iu_j$ in the basis in which the components the of $\textbf{u}_{i,\text{fit}}(j_n)$ and \smash{$\textbf{r}_{j_n(i)}^\text{Tem}$} are provided. The divergence can then be readily obtained as $u^0_{\text{fit}}=S_{11}+S_{22}$ (compression/expansion). We can also extract other strain components at position $\mathbf{r}_i$ from this matrix:  $u^1_{\text{fit}}=S_{21}-S_{12}$ (curl), $u^2_{\text{fit}}=S_{11}-S_{22}$ and $u^3_{\text{fit}}=S_{12}+S_{21}$ (shear components).

\subsubsection{Rotation frequency analysis}
\label{SI:rotfreq}
To measure embryo spinning frequencies within clusters, we took advantage of the inhomogeneous intensity profile within the embryo body that results from the uneven positioning of developing internal organs. In general, we find the rotation frequency that best rotates the embryo image from one frame to another, matching each embryo's inhomogeneous intensity profile. Specifically, for each embryo $i$ at time $t$, we first shift the origin of the coordinate system to the embryo centroid and apply its segmentation mask (section \ref{sec:SegmentTrack}) to select only the embryo body and reduce noise due to the intensity profile of other pixels surrounding it. After adjusting the pixel intensity of the masked image using the MATLAB function~\verb+imadjust+ to highlight the inhomogeneous intensity profile within the embryo, we obtain a saturated masked image of each embryo $i$, which we denote as $I_i(t)$. $I_i(t)$ can be rotated about its origin by angle $\varphi$ and then translated with vector $[\Delta x,\Delta y]$ using MATLAB functions ~\verb+imrotate+ and ~\verb+imtranslate+ with bicubic interpolation to obtain $I_i(t,\varphi,\Delta x,\Delta y)$. We further define a series of circular masks centered at the origin with radii equal to $0.4R_i$, $0.5R_i$, $0.6R_i$, $0.7R_i$, $0.8R_i$, where $R_i$ is the average radius of the segmentation mask of embryo $i$. We denote by function $F_j$ the application of circular mask number $j$ to an image. e.g., $F_1(I_i)$ returns the image $I_i$ with a circular mask of radius $0.4R_i$ applied. The rotation frequency of embryo $i$ at time $t$ is then obtained according to

\begin{equation}
    \omega_{ri}(t) = \left<\max_j\left\{\underset{\omega_r \in \Omega}{\arg\min}\left[\min_{[\Delta x,\Delta y]\in T}||F_j(I_i(t,0,0,0)-I_i(t+\Delta t,\omega_r \Delta t,\Delta x,\Delta y))||_2\right]\right\}\right>_{\Delta t},
\end{equation}
where $||\cdot||_2$ is the Euclidean norm of the matrix of pixel values. In other words, we find the rotation frequency $\omega_r$ that minimizes the squared error within circular mask $j$ between embryo images at time $t$ and $t+\Delta t$. In practice, we take $\Delta t$ to be $\pm 1$ frame. We maximize the rotation frequency found with different circular masks $j$ to eliminate the tendency of underestimation due to fixed background pixel intensity. To account for errors in centroid location measurements, we additionally minimize the squared error under translations of up to 3 pixels, denoted by $T$ in the above equation. Such a minimum is estimated by interpolating the squared error measured with discrete translations of $\pm 1.5$ and $\pm 3$ pixels using MATLAB function~\verb+fit+ with cubic spline interpolation. Similarly, the $\arg\min$ over $\omega_r$ is restricted in the range $\Omega = [-0.5,0.5]$ rad/frame, and estimated by interpolating the squared error measured with discrete $\omega_r$ increments of $0.05$ rad/frame using MATLAB function~\verb+fit+ with cubic spline interpolation. If for a particular $j$, the squared error has multiple minima as a function of $\omega_r$, the resulting $\omega_r$ measurement is considered unreliable and will be discarded.
Finally, to reduce noise in measured $\omega_{ri}(t)$ over time, we remove outliers defined as elements more than three local scaled median absolute deviation (MAD) from the local median over a time window of 10 frames using the MATLAB function~\verb|rmoutliers|, and then apply a Gaussian-weighted average over each window using the MATLAB function~\verb|smoothdata|. We run the analysis on the MIT SuperCloud \cite{reuther2018interactive} in parallel with 384 processes. In Fig. 4, embryos with discarded values of $\omega$ (NaN values) are outlined in black.

\subsubsection{Phase analysis of divergence strain and rotation frequency oscillations}
\label{SI:phase}
Following our model prediction that average divergence and rotation frequency oscillate in-phase in the oscillatory state (Fig. 4d-g in the main text), we wish to compare the phases of these two observables. To do so, we first assign their phases using a modified version of the method in \cite{Liu2021_topo}, which extracts the phase of a time-varying signal $I(t)$. This method works by first computing the moving center of the signal, $\bar{I}(t)$, using the MATLAB function \verb|movmean| with a time window approximately equal to the period of the oscillating time signal. Next we calculate the analytic extension of the centered signal $(I-\bar{I})(t)$ into the complex plane, $z(t) = (I-\bar{I})(t) + \mathbf{HT}[(I-\bar{I})(t)]$, where $\mathbf{HT}$ is the Hilbert transform computed using MATLAB function \verb|hilbert|. The phase of the signal is defined as the phase of $z(t)$.
After calculating the phases for the divergence and rotation frequency, we can straightforwardly plot them against each other to obtain the phase plots shown in the inset of Fig. 4 in the main text.

\subsection{Compression rate plot}
\label{SI:compression_plot}

To obtain Fig. 5c of the main text, we calculated the transverse velocity autocorrelations as a function of the lag time $\tau$, averaging over all embryos which persist throughout the entire duration of the experiment (indexed by $i$) and initial times $t_0$:
\begin{equation}
    \mathrm{VAC}_{\mathrm{trans}}(\tau) = \langle \mathbf{v}^i(t_0) \times \mathbf{v}^i(t_0+\tau) \rangle_{i,t_0} = \langle v_x^i(t_0) v_y^i(t_0+\tau) - v_x^i(t_0+\tau) v_y^i(t_0) \rangle_{i,t_0}
\end{equation}
Particle velocities $\mathbf{v}^i$ are first smoothed with a moving average of window width of 50 seconds to reduce noise. We choose to analyze transverse as opposed to longitudinal VACs to weight rotational and oscillatory motion more than (anti)parallel velocity signals.

\begin{figure}[h!]
  \centering
  \includegraphics[width=0.7\linewidth]{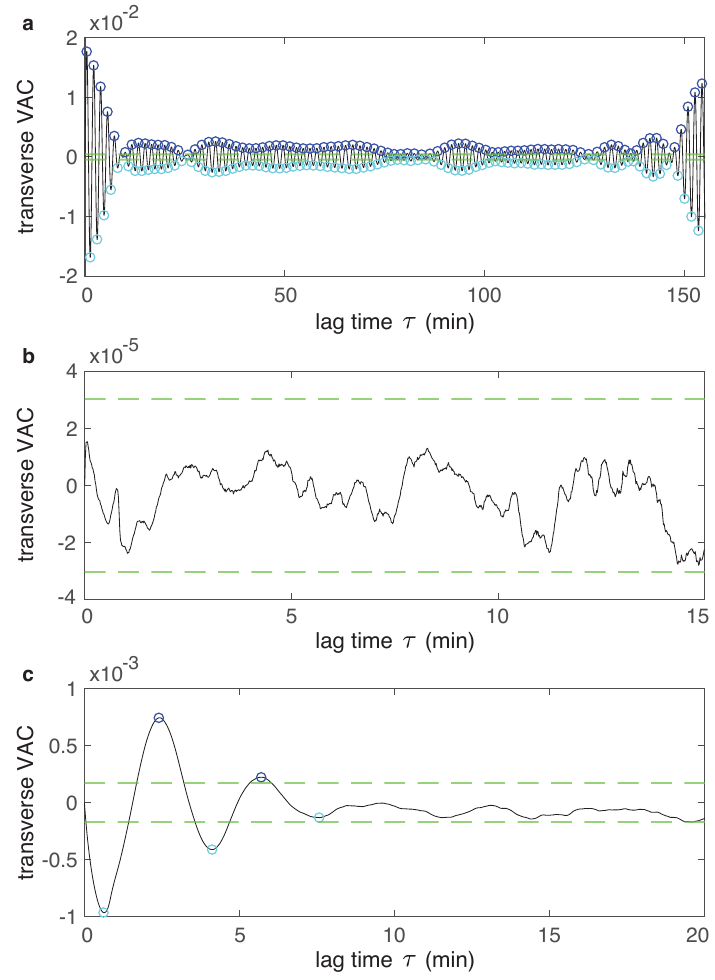}
    \caption{\textbf{Types of velocity autocorrelations in compression experiments.} Transverse velocity autocorrelation plots for mechanical compression-based experiments resulting in \textbf{(a)} persistent, \textbf{(b)} no, and \textbf{(c)} damped oscillations, as a function of lag time $\tau$. Blue (cyan) circles denote identified timelapse maxima (minima) used for calculation of oscillation duration. The noise threshold for minimum required peak-to-peak distance (between a minimum and maximum or vice-versa) is indicated by green dotted lines.}
  \label{fig:VAC_examples}
\end{figure}

We then identify all local extrema in this average signal using \texttt{MATLAB}'s \texttt{findpeaks} function, cutting off any with amplitudes less than or equal to a given threshold value $0.015\sqrt{\sigma_{\mathrm{VAC}}}$, where $\sigma_{\mathrm{VAC}}$ is the standard deviation of the velocity autocorrelation $\mathrm{VAC}(\tau)$. This noise threshold was determined by comparing analysis results with raw videos and choosing a value across all 77 mechanical compression experiments that consistently resulted in a number of oscillations that approximately agreed with what we observed by eye; avoided misidentification of smaller oscillation peaks as noise in cases where oscillation amplitudes shrink and then grow (as in fig. \ref{fig:VAC_examples}a); and allowed us to distinguish between noisy fluctuations (fig. \ref{fig:VAC_examples}b) and damped oscillations (see fig. \ref{fig:VAC_examples}c), the former of which tend to remain at a relatively constant amplitude.
We then calculated oscillation half-periods as the difference between the number of consecutive maxima/minima, and used twice the average value of this quantity to obtain the oscillation period. The oscillation duration was subsequently obtained as this oscillation period multiplied by the number of full oscillation periods.

To obtain the applied compression rate, we manually identified the time stamp and piston $y$-position at the beginning and in the end of compression for each step compression. We then normalized by the $y$ extent of the system prior to compression to obtain the average compression rate for each step compression.

\subsection{Irreversibility analysis}
\label{SI:irreversibility}
The irreversibility is calculated as the entropy production rate $\dot{S}$ of a system following overdamped Langevin dynamics, assuming the system is in steady state:
\begin{equation}
    \dot{S}=k_{\mathrm{B}}\int d\textbf{x}\,\frac{\mathbf{j}(\textbf{x})\cdot \mathbf{D}^{-1}(\textbf{x})\cdot \mathbf{j}(\textbf{x})}{\rho(\textbf{x})}.
    \label{eq:entropyproduction}
\end{equation}
Here, $k_{\mathrm B}$ is the Boltzmann constant. $\textbf{x}$ denotes some general system state, which will be replaced by the displacement field or different strain pairs in later analyses. $\rho(\textbf{x})$, $\mathbf{j}(\textbf{x})$, and $\mathbf{D}(\textbf{x})$ represent the corresponding state-dependent density, current, and diffusivity tensor, respectively.

For any given time series ${\textbf{x}_1, \textbf{x}_2, \dots, \textbf{x}_{N} }$ comprising a total of $N$ successive time points with intervals of $\Delta t$, we applied kernel estimation to construct the continuous density and associated current as follows:
\begin{subequations}\label{eq:EPdefs}
\begin{align}
    \hat{\rho}(\textbf{x})&=\frac{1}{N} \sum_{i=1}^{N} K(\textbf{x},\textbf{x}_i,\Sigma),
    \label{eq:density}\\
    \hat{\mathbf{j}}(\textbf{x})&=\frac{\hat{\rho}(\textbf{x})}{2\Delta t}
    \frac{\sum_{i=2}^{N-1} K(\textbf{x},\textbf{x}_i,\Sigma)(\textbf{x}_{i+1}-\textbf{x}_{i-1})}{\sum_{i=2}^{N-1} K(\textbf{x},\textbf{x}_i,\Sigma)},
    \label{eq:current}
\end{align}    
\end{subequations}
where $K(\textbf{x},\textbf{x}',\Sigma)=\exp[-(\textbf{x}-\textbf{x}')^\top\Sigma^{-1}(\textbf{x}-\textbf{x}')/2] /[2\pi\det(\Sigma)]^{-1/2}$ is the bivariate Gaussian with bandwidth $\Sigma_{\alpha \beta} = \frac{1}{2} N_t^{-1/6} \sigma_{\alpha} \delta_{\alpha \beta}$ determined using the so-called ``rule of thumb" \cite{li2019quantifying,bowman1997applied}, and $\sigma_{\alpha}$ is the standard deviation of observed variables $\textbf{x}$. 
Similarly, the diffusivity tensor is estimated as~\cite{just2003nonequilibrium}
\begin{equation}
    \bar{\mathbf{D}}(\textbf{x})=
    \frac{\hat{\rho}(\textbf{x})}{\Delta t}
    \frac{\sum_{i=1}^{N-1} K(\textbf{x},\textbf{x}_i,\Sigma)(\textbf{x}_{i+1}-\textbf{x}_{i})\otimes(\textbf{x}_{i+1}-\textbf{x}_{i})}{\sum_{i=1}^{N-1} K(\textbf{x},\textbf{x}_i,\Sigma)},
    \label{eq:difftens1}
\end{equation}
where $\otimes$ denotes a dyadic product. 
To ensure the existence of inversion, as required in eqn. \ref{eq:entropyproduction}, we use $\bar{\mathbf{D}}(\textbf{x})$ given in eqn. \ref{eq:difftens1} to define a constant, weighted mean diffusion matrix $\hat{\mathbf{D}}$ as
\begin{equation}
    \hat{\mathbf{D}}=
    \int d\textbf{x} \bar{\mathbf{D}}(\textbf{x})\hat{\rho}(\textbf{x}) {1}_{\hat{\rho}(\textbf{x}) > c}.
    \label{eq:diffusivetensor}
\end{equation}
Here, ${1}_{\hat{\rho}(\textbf{x}) > c}$ is the indicator function and $c$ is chosen such that only sufficiently well-populated regions of the phase space with $\hat{\rho}d\textbf{x}>0.01$ are included in the analysis. 
Equations \ref{eq:EPdefs} and \ref{eq:diffusivetensor} are then used in eqn. \ref{eq:entropyproduction} to approximate the irreversibility \smash{$\hat{\dot{S}}$}. 

\begin{figure}[h!]
  \centering
  \includegraphics[width=0.6\linewidth]{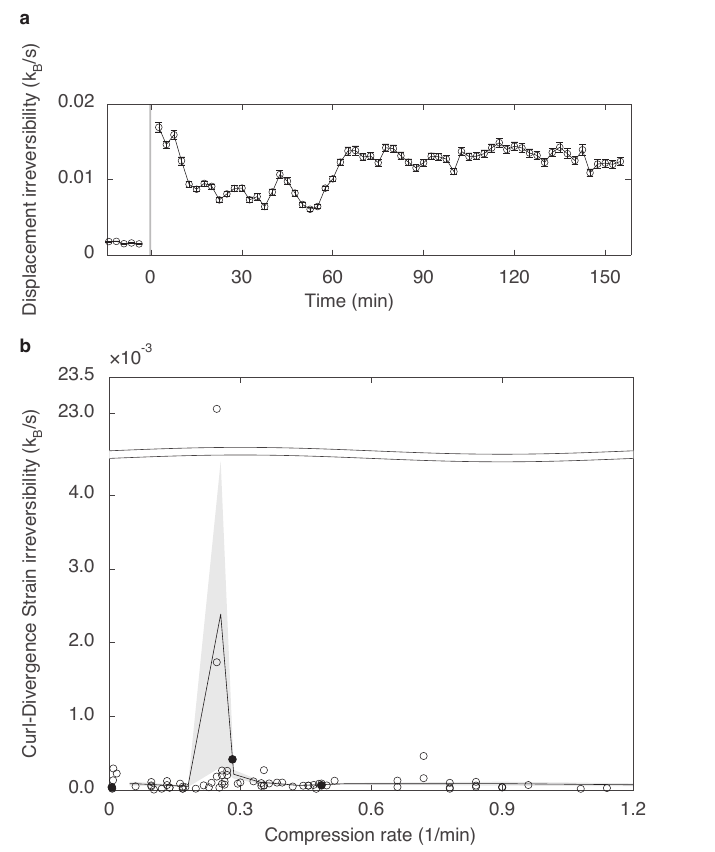}
    \caption{\textbf{Irreversibility of compression-induced oscillations.} \textbf{(a)} Displacement irreversibility of a representative experiment before and after step compression, at an intermediate compression rate inducing long-lived oscillations. The shaded gray bar around 0 min represents the compression in progress. The measured irreversibility is significantly higher after the compression step. \textbf{(b)} Irreversibility of curl-divergence strain as a function of applied compression rate. Line and shading represent mean and standard error of the mean for binned compression rate. The three filled circles correspond to the representative experiments shown in Fig. 5.}
  \label{fig:CompOscIrr}
\end{figure}

To calculate the displacement irreversibility, we first divide the smoothed displacement field (section \ref{sec:smDisp}) into square regions with side length of approximately $200 \mu$m.
The center of each square is positioned at some position $\textbf{r}$. 
Within each of these squares, we spatially average the displacement field and introduce the local displacement field $\hat{\textbf{u}}(\textbf{r},t)$. 
Replacing our general system state $\textbf{x}$ with $\hat{\textbf{u}}(\textbf{r})$, we can construct the local irreversibility $\hat{\dot{S}}(\textbf{r})$.
Each local irreversibility approximates the entropy production rate of the embryo at $\textbf{r}$ from its displacements. 
Finally, we spatially integrate the local irreversibility for the whole cluster
\begin{equation}
    \hat{\dot{S}}_{\mathrm tot} = \sum_{\textbf{r}} \hat{\dot{S}}(\textbf{r}), 
\end{equation}
to estimate the total irreversibility. 
The time series of the estimated total irreversibility before and after compression for an exemplary experiment compressed at an intermediate applied strain rate is plotted in fig. \ref{fig:CompOscIrr}a. In fig. \ref{fig:CompOscIrr}b, each dot represents the total curl-divergence strain irreversibility of an experiment compressed at the specified applied strain rate, integrated over all time post-compression. The corresponding displacement irreversibility is plotted in Fig. 5c (bottom).

To calculate the strain irreversibility, the strain field is directly derived from the smoothed displacement field and binned in 200 $\mu m$ squares. 
Local strain components $u^{\alpha}$ are then spatially averaged within each square, and we consider the strain component pair $[u^{\alpha}(\textbf{r},t), u^{\beta}(\textbf{r},t)]$ for our irreversibility analysis. 
Inspired by the curl-divergence and shear1-shear2 coupling of odd elastic materials, our analysis focuses on pairs $\alpha=1$, $\beta=0$ (fig. \ref{fig:CompOscIrr}b).

\subsection{Neighbor exchange rate}
\label{SI:neighborexchange}
\begin{figure}[h!] 
  \centering
  \includegraphics[width=0.7\linewidth]{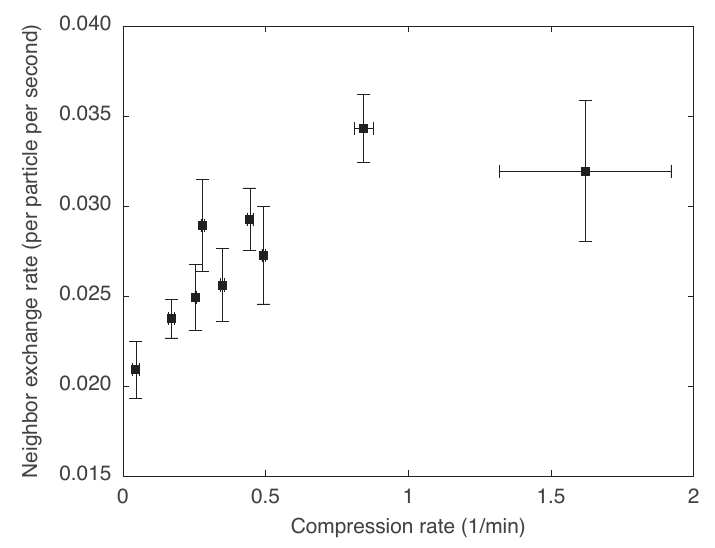}
    \caption{\textbf{Rate of neighbor exchange in mechanical compression experiments.} The rate at which the local neighborhood of a given embryo changes tends to increase with applied compression rate, corresponding to an increase in plastic deformations.}
  \label{fig:NNexchangeStrainrate}
\end{figure}

To quantify neighbor exchange rate after step compression, we first determine each embryo's nearest neighbors at each time point using a distance cutoff. Specifically, the nearest neighbors of embryo $i$ are defined as those embryos that are separated by a distance less than 1.3 times the lattice constant from embryo $i$, where the lattice constant is measured as the location of the first peak of the radial distribution function $g(r)$. Only embryos with at least 5 nearest neighbors are considered in this analysis. For each embryo, we compare the list of nearest neighbors between two consecutive frames to count the number of times that either: the identity of one or more nearest neighbors changes; or, the number of nearest neighbors changes for each embryo. As shown in fig. \ref{fig:NNexchangeStrainrate}, the neighbor exchange rate (per embryo per second) tends to increase as a function of applied compression rate. 

\section{Simulations}
\label{SI:simulations}
\begin{figure}[h!]
  \centering
\includegraphics[width=0.95\linewidth]{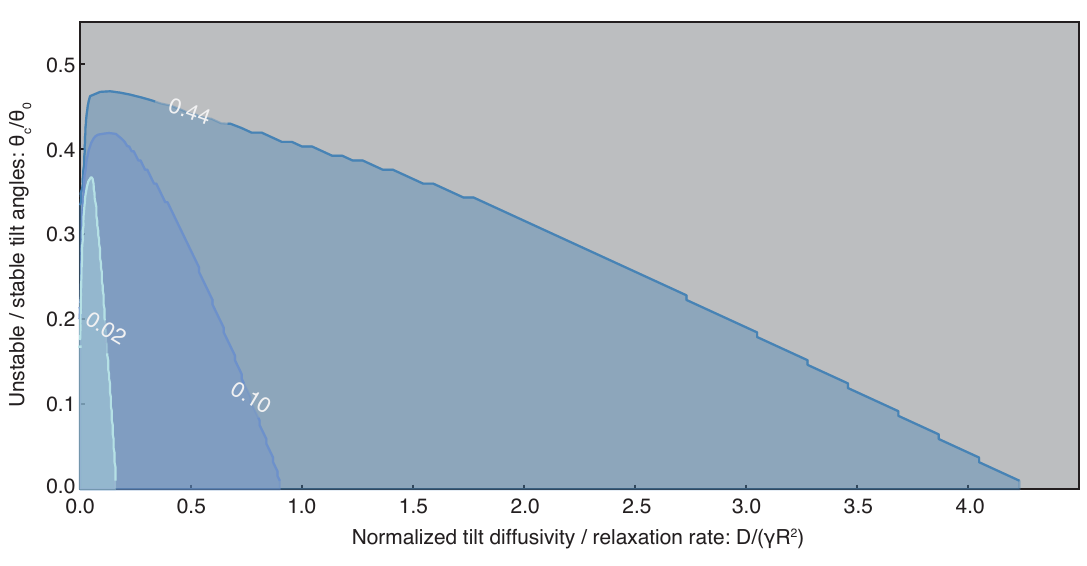}
    \caption{\textbf{Numerical phase diagram in active wave regime of odd elasticity for small tilt-displacement coupling.} Range expansion phase diagram for tilt-induced oscillations in the no-wave regime of odd elasticity with $(B,\mu,A,K_0) = (7, 3.5, 27, 25)$ and tilt-displacement coupling $\beta = 0.1$, as a function of (un)stable tilt angle ($\theta_c$) $\theta_0$, tilt diffusion constant $D$, and relaxation rate $\gamma$. Regions in which steady-state tilt oscillations are permitted are colored in shades of blue and labeled with the area fraction of initial $\theta>0$ (see main text). Regions where oscillations are not permitted are colored in gray. }
  \label{fig:oddwave_PD}
\end{figure}

\begin{figure}[h!]
  \centering
  \includegraphics[width=0.95\linewidth]{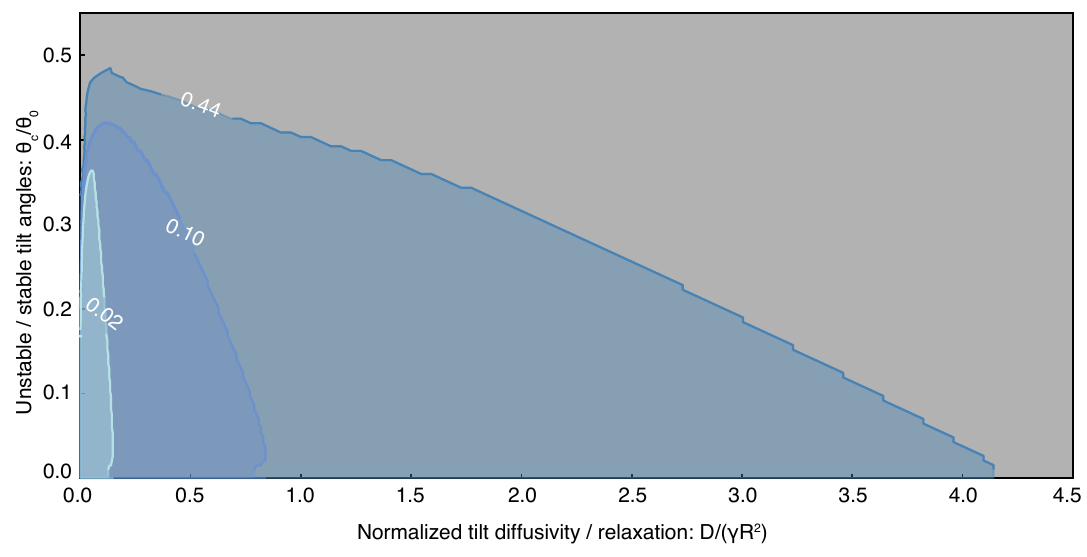}
    \caption{\textbf{Numerical phase diagram in no-wave regime of odd elasticity for small tilt-displacement coupling.} Range expansion phase diagram for tilt-induced oscillations in the no-wave regime of odd elasticity with $(B,\mu,A,K_0) = (7, 3.5, 1.9, 0.8)$ and tilt-displacement coupling $\beta = 0.1$, as a function of (un)stable tilt angle ($\theta_c$) $\theta_0$, tilt diffusion constant $D$, and relaxation rate $\gamma$. Regions in which steady-state tilt oscillations are permitted are colored in shades of blue and labeled with the area fraction of initial $\theta>0$. Regions where oscillations are not permitted are colored in gray. }
  \label{fig:nowave_PD}
\end{figure}

\begin{figure}[h!]
  \centering
  \includegraphics[width=0.95\linewidth]{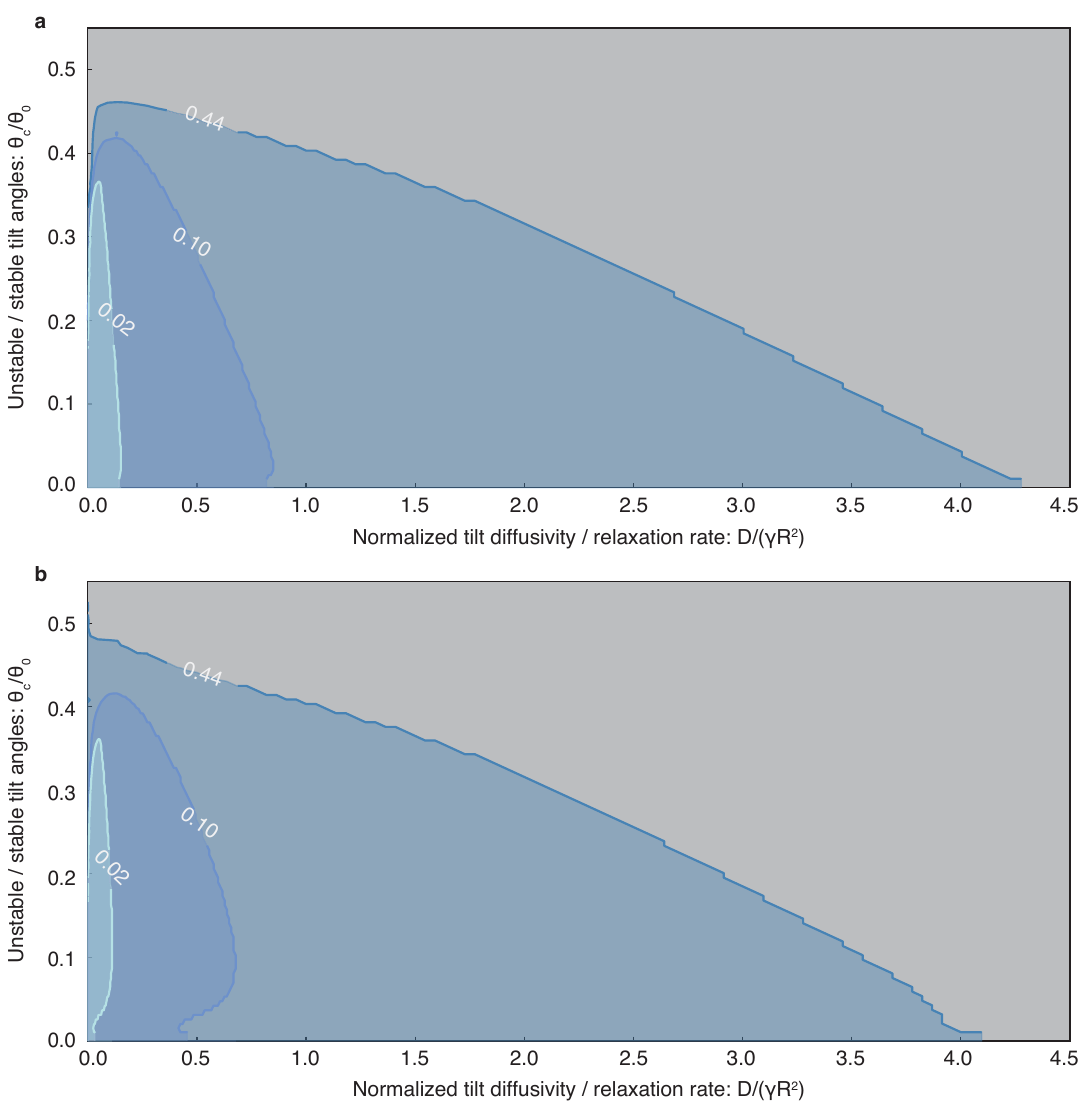}
    \caption{\textbf{Numerical phase diagrams for large tilt-displacement coupling.} Range expansion phase diagram for tilt-induced oscillations in the \textbf{(a)} active wave and \textbf{(b)} no-wave regimes of odd elasticity for tilt-displacement coupling $\beta = 1$ (see section S2.1), corresponding to $(B,\mu,A,K_0) = (7, 3.5, 27, 25)$ and $(B,\mu,A,K_0) = (7, 3.5, 1.9, 0.8)$, respectively. Regions in which steady-state tilt oscillations are permitted are colored in shades of blue and labeled with the area fraction of initial $\theta>0$. Regions where oscillations are not permitted are colored in gray. }
  \label{fig:beta1_PDs}
\end{figure}

We use the \texttt{Dedalus} framework to simulate the model given in eqns. \ref{eq:S1}-\ref{eq:S4} \cite{dedalus}. For each point on the phase diagrams in fig. \ref{fig:oddwave_PD}-\ref{fig:beta1_PDs}, we define a 100 x 100 grid with periodic boundary conditions in both directions. 

To simulate the stability of oscillatory and fluctuating stable states in different parameter regimes, we initialize a small oscillating patch of nonzero tilt angles $\theta$ in the center of the system, setting the remaining grid points to be in the fluctuating state, i.e. with $\theta_0=\theta(t=0) = 0$. Displacements $\textbf{u}$ and orientations $\phi$ are initialized with randomized fields pulled from a uniform distribution. 

As discussed in section \ref{sec2:4}, the growth / relaxation and diffusion terms in eqn. \ref{eq:S4} allow for spatial expansion and contraction of the oscillatory patch. Since our model is deterministic, if the oscillatory state is not stable for a given choice of parameters, the system will eventually equilibrate at $\theta = 0$ uniformly. We as such run each gridpoint simulation up to a final simulation time $t_f=300$ with a simulation timestep of $\Delta t = 0.005$ for three different area fractions of the initial patch of nonzero $\theta$. From each simulation, we define the fraction of the system oscillating as the proportion of gridpoints with $\theta(t=t_f) > 10^{-5}$. For systems with an average speed $\langle |\mathbf v| \rangle < 10^{-6}$, we set this oscillating fraction equal to 0 to avoid situations in which the $\theta$ field might be some large nonzero value even as the displacements themselves have been completely suppressed (by diagonal terms in the elastic modulus tensor $C_{ijmn}$). 

To extract the phase boundary between oscillatory and fluctuating states, we extract the level contour at which the oscillating fraction equals 0.9 for a given initial $\theta > 0$ area fraction. Overlaying these contours for elastic moduli $(B,\mu,A,K_0) = (7, 3.5, 27, 25)$ in the active strain wave regime of odd elasticity \cite{scheibner2020odd} results in fig. \ref{fig:oddwave_PD}. On the other hand, even choosing parameters in a parameter regime not producing odd elastic waves does not result in a qualitatively different phase diagram, as shown in fig. \ref{fig:nowave_PD}, for $(B,\mu,A,K_0) = (7, 3.5, 1.9, 0.8)$. In both cases, we take tilt-displacement coupling $\beta = 0.1$ to be small, assuming the range expansion of the experimental oscillations to dominate tilt angle ($\theta$) dynamics over strain contributions. Large $\beta$ serves to shrink the region of allowed oscillation, to a greater extent in systems without odd elastic wave propagation - compare fig. \ref{fig:oddwave_PD}, fig. \ref{fig:nowave_PD} and fig. \ref{fig:beta1_PDs}b. This is as expected, as overdamped systems without odd waves but large tilt-displacement coupling $\beta$ will cause tilt dynamics to damp out more quickly. However, this effect is prevalent largely only at small unstable tilt angles $\theta_c$; varying $\beta$ in general does not otherwise appear to substantially alter the regions of phase space in which sustained oscillations can be achieved. 

Full implementation of eqns. \ref{eq:S1} - \ref{eq:S4} require quantification of various model parameters such as the torque exchange term $\tau_0 = 0.12$, which we obtain from microscopic estimates in previous work \cite{tan2022odd}. For the rotational drag coefficients $\Lambda_{AP,\perp}$ around the embryo's anterior-posterior axis and the axes perpendicular to it, respectively, introduced in eqn. \ref{eq:S3}, we assume the rotational drag that the embryos experience is proportional to the torque on a spinning sphere in a fluid. For a sphere with radius $R$ in a fluid of viscosity $\eta$, this drag term is given by $f_{sphere} = 8\pi \eta R^3$, where $R$ is given by the length of the embryo perpendicular to the axis of rotation, e.g. $m/2$ when rotating about the AP axis. Then taking $M_{AP}$ and $m$ as an embryo's AP and perpendicular (minor) axis lengths, respectively, we have $\Lambda_\perp \sim M_{AP}^3$ and $\Lambda_{AP} \sim m^3$, resulting in 
\begin{equation}
    \frac{\partial \phi}{\partial t} = \frac{\sin(k\theta)\omega_s}{\left[\alpha^3 \sin\theta\cos(k\theta) - \cos\theta\sin(k\theta)\right]} \left(  1+\frac{6Rc}{d_0}\partial_k u_k \right)
\end{equation}
where $\alpha = M/m$ is the aspect ratio of an embryo. We use $\alpha = 1.4$, the 94th percentile of aspect ratios in the experiment described in the main text. Since we know that $k\theta$ and $\theta$ must both be less than the stable tilt angle $\theta_0$, our constraint on $k$ is that $k \leq 1$. We assume the embryos' average tilt angle $\theta$ and the angle of mismatch between active torque and force $\theta_M$ are close together, and take $k=0.9$ (see section S\ref{sec2:2}). To generate the phase diagrams shown in this section and in the main text, we run batches of simulations in parallel on the MIT SuperCloud \cite{reuther2018interactive}.

\clearpage
\section{Table of symbols}
{\centering \footnotesize
\begin{longtable}{| p{.25\textwidth} | p{.75\textwidth} |} 
\hline
$\mathbf{u}_i$ & Instantaneous displacement of embryo $i$\\
$\mathbf{v}_i$ & Instantaneous velocity of embryo $i$\\
$u_{x,y}, u_{ij} = \partial_i u_j$ & Displacement field components and displacement gradient tensor\\
$u^\alpha$ & Strain components: divergence (compression/expansion; 0), curl (rotation; 1), and shear (2, 3) \\
$\omega_{ri}$ & Angular spinning frequency of disk/particle $i$\\
$d_{ij}$ & Shortest disk surface distance between disks $i$ and $j$\\
$v_0$ & Continuum theory maximal self-propulsion speed \\
$\theta$ & Coarse-grained tilt angle in continuum theory \\
$\phi$ & Planar orientation angle in continuum theory dictating direction of self-propulsion \\
$\omega_{s,p}$ & Coarse-grained spin and precession frequencies \\
$\eta$ & Fluid viscosity\\
$\mathbf{r}=(x,y,z)^\top$ & Position vector with magnitude $r=|\mathbf{r}|$\\
$C_{ijmn}$ & Elastic modulus tensor, containing both odd and even components \\ 
$B$, $\mu$ & Bulk modulus, Shear modulus\\
$A$, $K_0$ & Odd bulk modulus, Odd shear modulus\\
$\Lambda_{AP}$, $\Lambda_\perp$ & Rotational drag coefficients \\
$R$ & Apparent embryo top-down radius \\
$d_0$ & equilibrium surface to surface distance between embryos \\
$d_{ij} = |\mathbf{r}_i - \mathbf{r}_j| -2R$ & Surface-to-surface distance between embryos $i$ and $j$ \\
$\gamma$ & Tilt angle growth/relaxation rate \\
$\theta_c,\theta_0$ & Unstable and stable tilt angles \\
$D$ & Tilt angle diffusion coefficient \\
$\beta$ & Tilt-displacement coupling strength \\
$\vec F_a$, $\vec \tau_a$ & Active force and torque generated by an embryo \\
$\mathbf \Omega$ & Angular velocity vector ($|| \vec tau_a$) \\
$\varphi,\phi,\psi$ & Euler angles \\
$\theta_M$ & angle of mismatch between $\vec F_a$ and $\vec \tau_a$ \\
$d_c$ & Range of torque exchange interactions \\
$\tau_0$ & Torque exchange interaction strength \\
$A(q_0)$ & Gaussian kernel amplitude associated with inverse lengthscale $q_0$, used to model compression step \\
$\mathcal L(x_0,y_0,\omega,c_x,c_y)$ & Loss function for rigid body fit \\
$c_{x,y}$ & Center of global rotation, for calculating corotating frame \\
$x_0,y_0$ & Averaged initial positions of embryos\\
$\hat \rho(\mathbf x)$ & Continuum probability density \\
$\hat j(\mathbf x)$ & Probability current \\
$M_{AP}$ & Embryo anterior-posterior (AP) body-axis length\\
$M_{proj}$ & Projection of $M_{AP}$ onto the xy (imaging) plane \\
$m$ & Embryo minor axis length (assuming ellipsoidal shape) \\
$\alpha=M_{AP}/m$ & Embryo aspect ratio in the xy-plane \\
$N_{cl}$ & Total number of embryos or disks bound in a cluster \\
$\phi_n$ & Dynamical Mode Decomposition (DMD) eigenvectors (length 2$N$) \\
$-\lambda_n + i\nu_n$ & DMD eigenvalues (real and imaginary components)\\
$\mathbf u_n(t) = \phi_n e^{(i\omega_n-\lambda )t}$, $\mathbf u_{n,\mathcal R}$, $\mathbf u_{n,\mathcal I}$ & DMD mode number $n$ and associated real/imaginary components \\
$\dot S$ & Entropy production rate \\
$C_{n,m}^{\mathcal R,\mathcal R}$, $C_{n,m}^{\mathcal I,\mathcal R}$, $C_{n,m}^{\mathcal I,\mathcal I}$ & Normalized real-real, imaginary-real, and imaginary-imaginary correlation matrices \\
$\mathcal L_{n,m}^{\mathcal R, \mathcal R}(\tau)$ & Angular momentum tensor (antisymmetrized cross-correlation) \\
\hline
\end{longtable}

}

\newpage
\bibliographystyle{unsrt}
\bibliography{references}